\documentclass[twocolumn,floatfix,superscriptaddress,showpacs,showkeys]{revtex4}
\usepackage{amsmath}
\usepackage{graphicx}
\usepackage{dcolumn}
\usepackage{bm}
\usepackage{booktabs}
\usepackage{amssymb}
\usepackage{multirow}
\usepackage{makecell}
\usepackage{textcomp}
\usepackage{subfigure}
\usepackage{phonetic}
\usepackage{extarrows}
\usepackage{color}
\usepackage{float}
\usepackage[colorlinks,citecolor=blue,linkcolor=blue]{hyperref}

\begin{document}


\title{Dissipation-induced topological phase transition and periodic-driving-induced photonic topological state transfer in a small optomechanical lattice}

\author{Lu Qi}
\affiliation{School of Physics, Harbin Institute of Technology, Harbin, Heilongjiang 150001, China}
\author{Guo-Li Wang}
\affiliation{School of Physics, Harbin Institute of Technology, Harbin, Heilongjiang 150001, China}
\author{Shutian Liu}
\email{stliu@hit.edu.cn}
\affiliation{School of Physics, Harbin Institute of Technology, Harbin, Heilongjiang 150001, China}
\author{Shou Zhang}
\email{szhang@ybu.edu.cn}
\affiliation{School of Physics, Harbin Institute of Technology, Harbin, Heilongjiang 150001, China}
\affiliation{Department of Physics, College of Science, Yanbian University, Yanji, Jilin 133002, China}
\author{Hong-Fu Wang}
\email{hfwang@ybu.edu.cn}
\affiliation{Department of Physics, College of Science, Yanbian University, Yanji, Jilin 133002, China}


\date{\today}

\begin{abstract}
We propose a scheme to investigate the topological phase transition and the topological state transfer based on the small optomechanical lattice under the realistic parameters regime. We find that the optomechanical lattice can be equivalent to a topologically nontrivial Su-Schrieffer-Heeger (SSH) model via designing the effective optomechanical coupling. Especially, the optomechanical lattice experiences the phase transition between topologically nontrivial SSH phase and topologically trivial SSH phase by controlling the decay of the cavity field and the optomechanical coupling. We stress that the topological phase transition is mainly induced by the decay of the cavity field, which is counter-intuitive since the dissipation is usually detrimental to the system. Also, we investigate the photonic state transfer between the two cavity fields via the topologically protected edge channel based on the small optomechanical lattice. We find that the quantum state transfer assisted by the topological zero energy mode can be achieved via implying the external lasers with the periodical driving amplitudes into the cavity fields. Our scheme provides the fundamental and the insightful explanations toward the mapping of the photonic topological insulator based on the micro-nano optomechanical quantum optical platform.     	
\end{abstract}

\maketitle


\section{\label{sec.1}Introduction}
Recently, topological insulator~\cite{hasan2010colloquium,qi2011topological,chiu2016classification,bansil2016colloquium} has attracted widespread attention in the field of condensed matter physics since it possesses numerous novel properties, such as the commensal existences of the insulating bulk state and the conducting edge state. These newfangled properties lead topological insulator have abundant potential applications in quantum information processing~\cite{dlaska2017robust,mei2018robust,qi2020controllable}, quantum computing~\cite{aasen2016milestones,Sarma2015}, topological laser~\cite{harari2018topological,bandres2018topological}, etc. The previous investigations in the field of topological insulator mainly aim to the solid state electronic system~\cite{hasan2010colloquium,qi2011topological,bansil2016colloquium}, in which the topological models of the single-electron or multi-electron have been widely investigated. With the very fast developing of the micro-nano processing technology and the quantum optical devices, multifarious bosonic optical systems have became the excellent-performance and very-promising platforms for the investigations of photonic topological insulator~\cite{khanikaev2013photonic,rechtsman2013photonic}, including the topological phase and phase transition in photonic crystal~\cite{yan2014topological,lu2016symmetry,gao2015stable,shalaev2019robust,skirlo2015experimental,xu2016accidental,wu2015scheme}, photonic topological insulator and photonic topological invariant in optical fiber and optical waveguide~\cite{tomita1986observation,chen2014experimental,ke2019topological,longhi2013zak,blanco2016topological}, the analog of the topological insulator in optical cavity and optical resonator~\cite{liang2013optical,mivehvar2017superradiant,liang2014optical,poli2015selective,he2016topological,leykam2018reconfigurable,tangpanitanon2016topological}, the topological edge states and topological insulator in silicon photonics~\cite{hafezi2013imaging,ma2016all,he2019silicon}, topological semimetal phase and topological phase transition in circuit quantum electrodynamics system~\cite{mei2016witnessing,huang2016realizing,mei2015simulation,qi2018simulation,tan2019simulation}, etc.

In addition to the optical systems mentioned above, the optomechanical system~\cite{aspelmeyer2014cavity,kippenberg2008cavity,eichenfield2009optomechanical,kippenberg2007cavity} consisting of the radiation-pressure-coupled optical cavity field and the micro-mechanical resonator, as a kind of burgeoning quantum optical system, has drawn growing attention for the investigations of various quantum matters both in macro and micro scales~\cite{wang2018normal,dobrindt2008parametric,liu2013dynamic,vitali2007optomechanical,ghobadi2014optomechanical,bai2019engineering,purdy2013strong,bai2019qubit,nunnenkamp2010cooling,wang2018optomechanical}. Analogously, the optomechanical lattice, which is composed by multiple sets of single optomechanical system, has also became an excellent candidate to handle the diversified multi-body physics issues, such as the collective dynamics in optomechanical array~\cite{heinrich2011collective,ludwig2013quantum,xuereb2012strong}, the photon-phonon entanglement in coupled multi-optomechanical system~\cite{akram2012photon}, the long-range optical transmission~\cite{xiong2015asymmetric}, the dynamical phase transitions and excitation of solitons in optomechanical lattice~\cite{qi2019bosonic,tomadin2012reservoir,gan2016solitons}, etc. Furthermore, the multiple cavity fields and the resonators of the optomechanical lattice, under the steady-state dynamics, provide us a physical opportunity for the mapping of the topological tight-binding model. For example, in Ref.~\cite{qi2017simulating}, a scheme has been proposed to induce a $Z_{2}$ topological insulator based on a optomechanical array. In Ref.~\cite{Roque2017Anderson}, the Anderson localization effect has been revealed in a disordered optomechanical arrays. Especially, in Ref.~\cite{Raeisi2020Quench}, a topologically nontrivial Su-Schrieffer-Heeger (SSH) model has been mapped based on the optomechanical arrays with quench dynamics. However, the effects of the inherent dissipation on the topology of the system, especially on topological phase transition, in all of the above schemes, are not yet clearly revealed.                

In this paper, inspired by the matter mentioned above, we propose a scheme to investigate the topological phase and the phase transition based on a small optomechanical lattice. We reveal that the effective steady-state Hamiltonian of the optomechanical lattice can be mapped into a tight-binding Hamiltonian only possessing the nearest-neighboring hopping. We find that, via designing the effective optomechanical coupling, the effective tight-binding Hamiltonian can be equivalent to a SSH model with the nontrivial topology. Meanwhile, we propose two different ways to resist the adverse impacts of the finite-size effect caused by small lattice on the topology of the system. Dramatically, by controlling the decay of the cavity field and the optomechanical coupling, the topologically nontrivial SSH phase can be transformed into a topologically trivial SSH phase. In this way, the small optomechanical lattice experiences the phase transition between topologically nontrivial SSH phase and topologically trivial SSH phase. Further, we investigate the photonic state transfer of the cavity fields assisted by the topological edge channel when the periodical drivings  are added into the cavity fields. The numerical results reveal that the photonic state transfer can be achieved based on the topological zero energy mode in the gap by choosing  the periodical driving appropriately. We stress that our scheme explicates the feasibility and the rationality of inducing the topological phase transition via the inherent dissipation, which provides more possibilities of the mapping between the optomechanical system and the photonic topological insulator community.   

The paper is organized as follows: In Sec.~\ref{sec.2}, we derivate the effective steady-state Hamiltonian of the small optomechanical lattice and realize the mapping of the tight-binding Hamiltonian. In Sec.~\ref{sec.3}, we focus on the topological SSH phase transition induced by the decay of the cavity field and the optomechanical coupling. In Sec.~\ref{sec.4}, we investigate the photonic topological state transfer via implementing the periodical drivings of the cavity fields. Finally, a conclusion is given in Sec.~\ref{sec.5}.

\section{\label{sec.2} Model and Hamiltonian}
\begin{figure}
	\centering
	\includegraphics[width=0.9\linewidth]{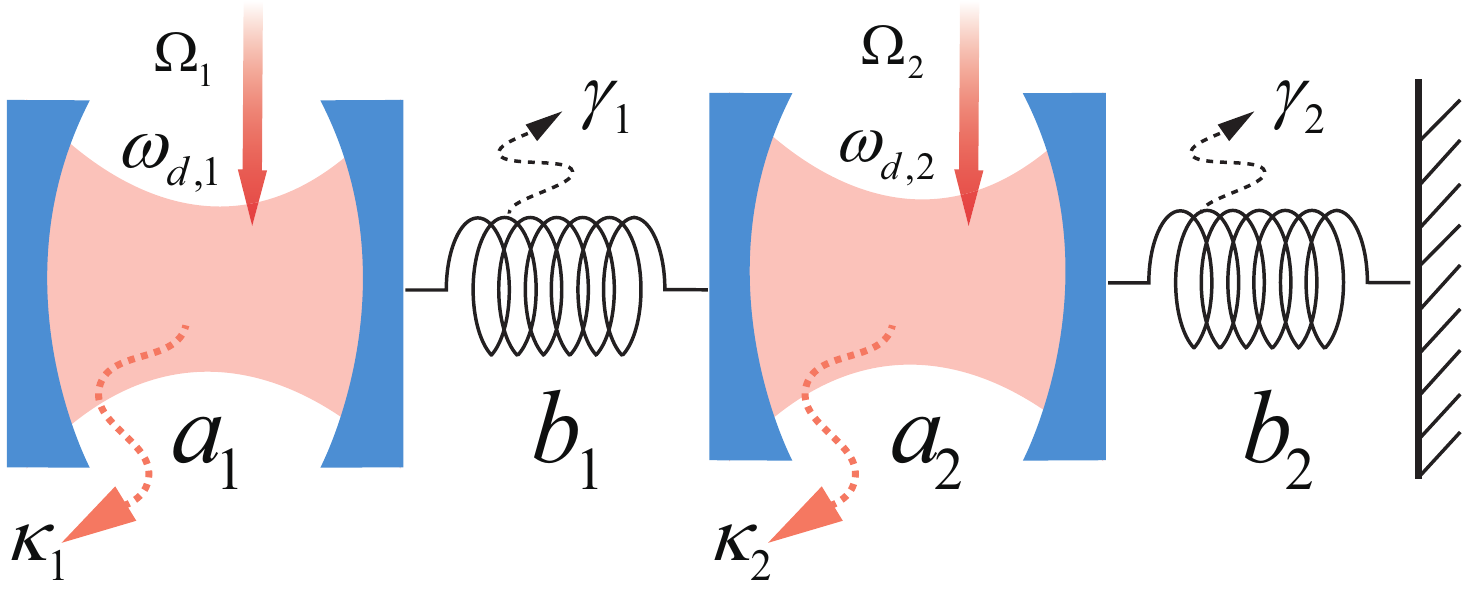}\\
	\caption{Schematic of the 1D small optomechanical lattice. The optomechanical lattice contains two unit cells, in which each unit cell consists of a cavity field and a resonator. The cavity field $a_{1}$ ($a_{2}$) is driven by an external laser with driving amplitude $\Omega_{1}$ ($\Omega_{2}$) and driving frequency $\omega_{d,1}$ ($\omega_{d,2}$). The decay of the cavity field $a_{1}$ ($a_{2}$) and the damping of the resonator $b_{1}$ ($b_{2}$) are $\kappa_{1}$ ($\kappa_{2}$) and $\gamma_{1}$ ($\gamma_{2}$), respectively.}\label{fig1}
\end{figure}
\begin{figure*}
	\centering
	\subfigure{\includegraphics[width=0.32\linewidth]{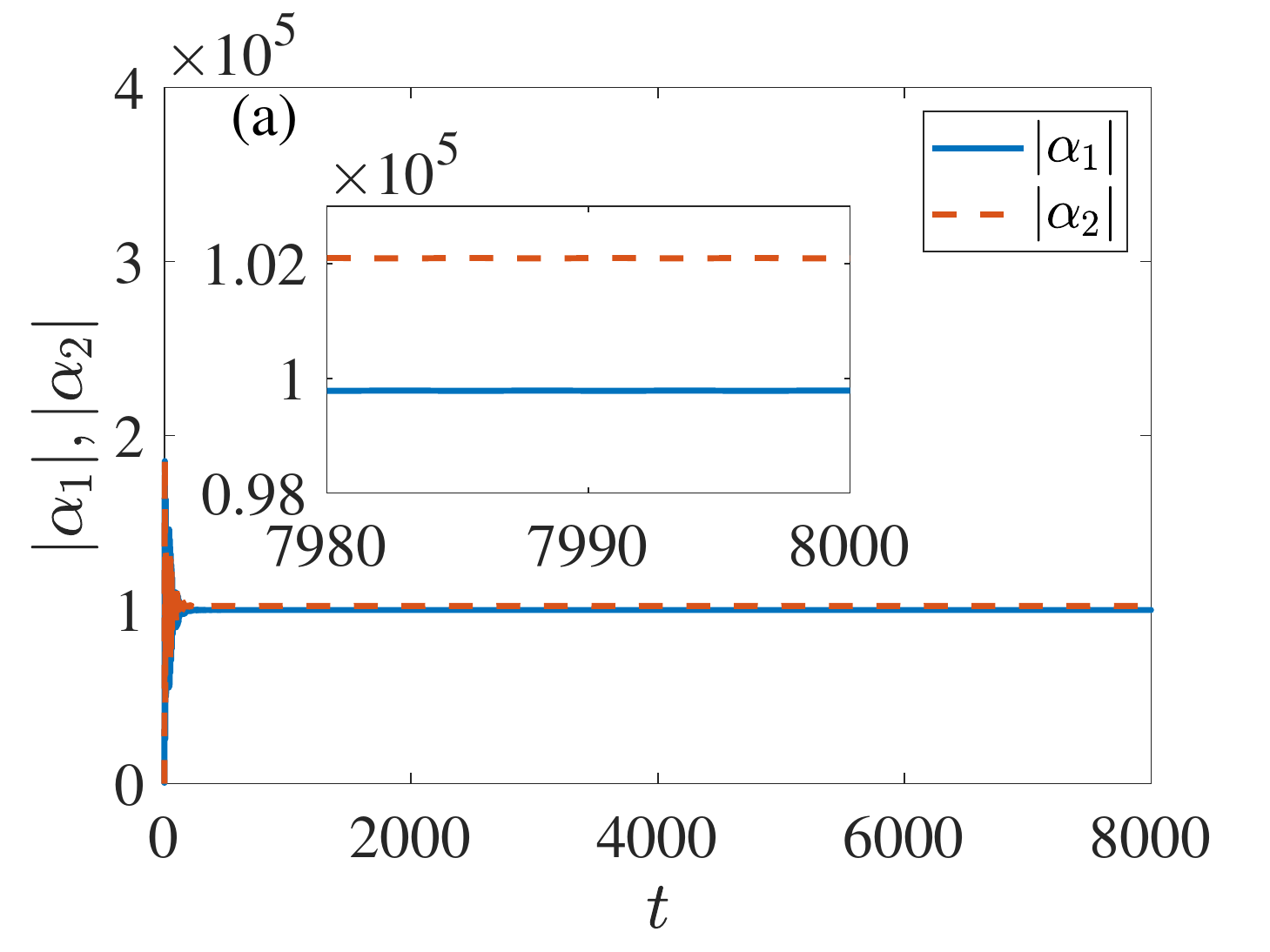}}
	\subfigure{\includegraphics[width=0.32\linewidth]{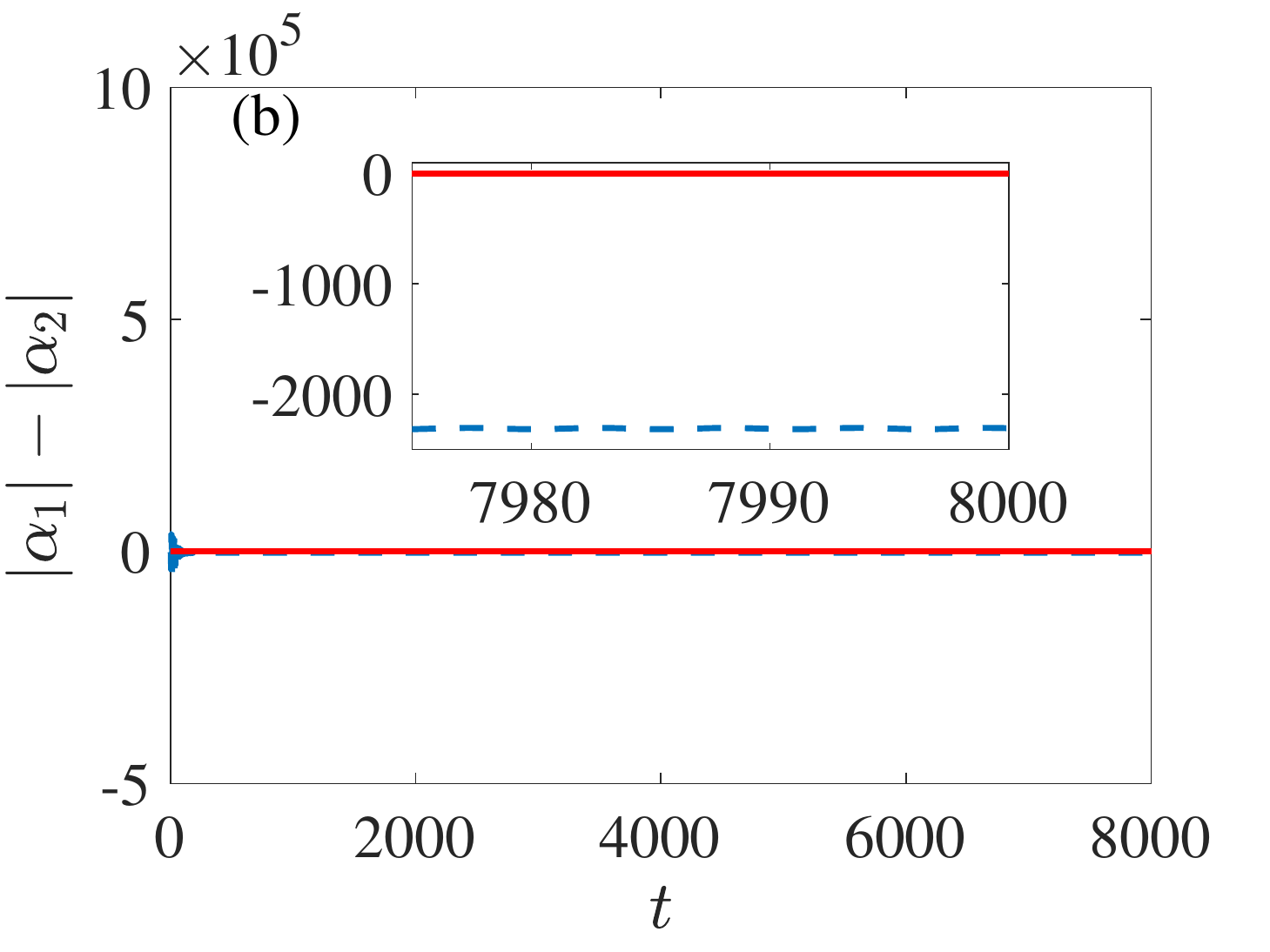}}
	\subfigure{\includegraphics[width=0.32\linewidth]{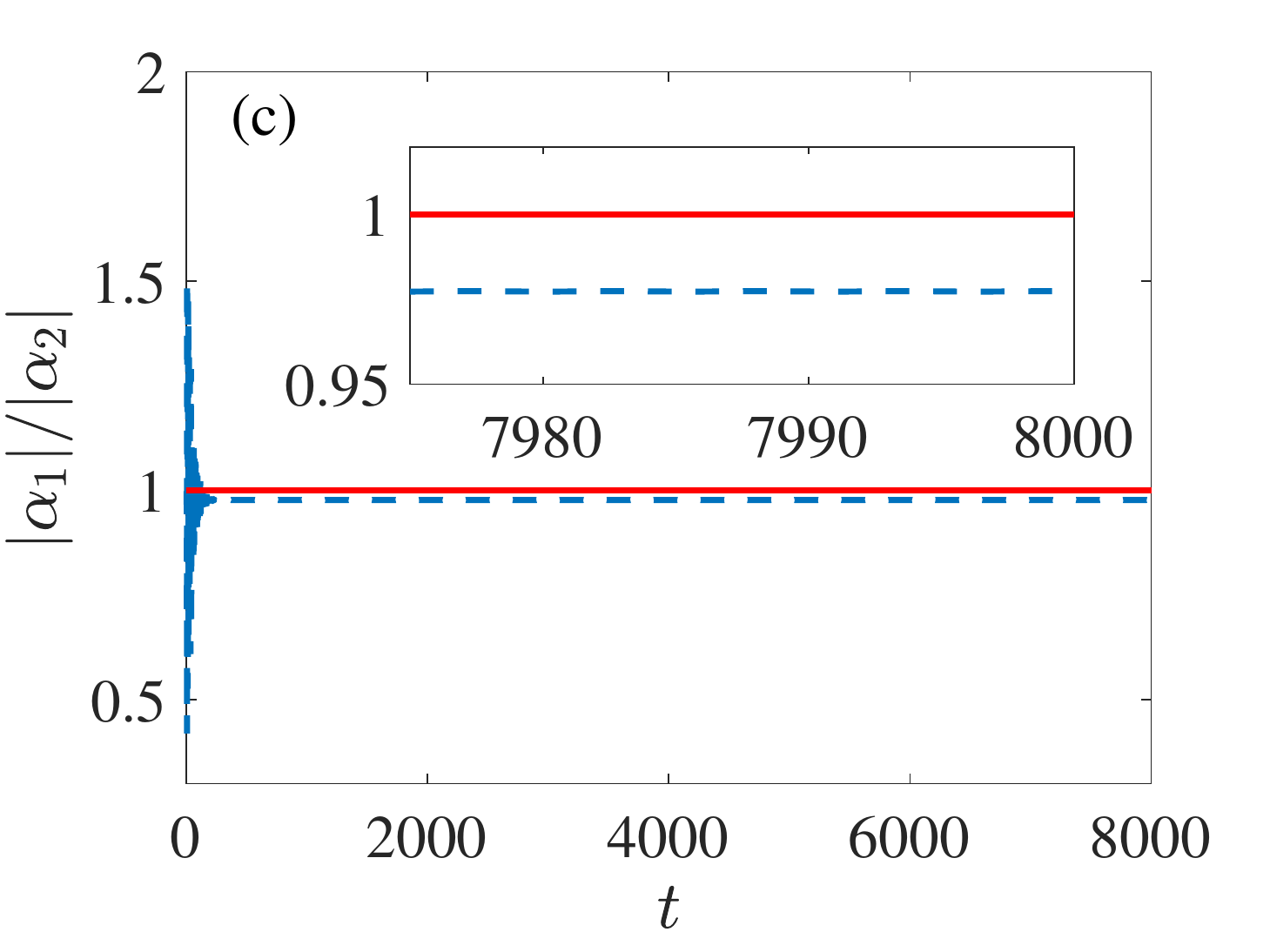}}
	
	\subfigure{\includegraphics[width=0.32\linewidth]{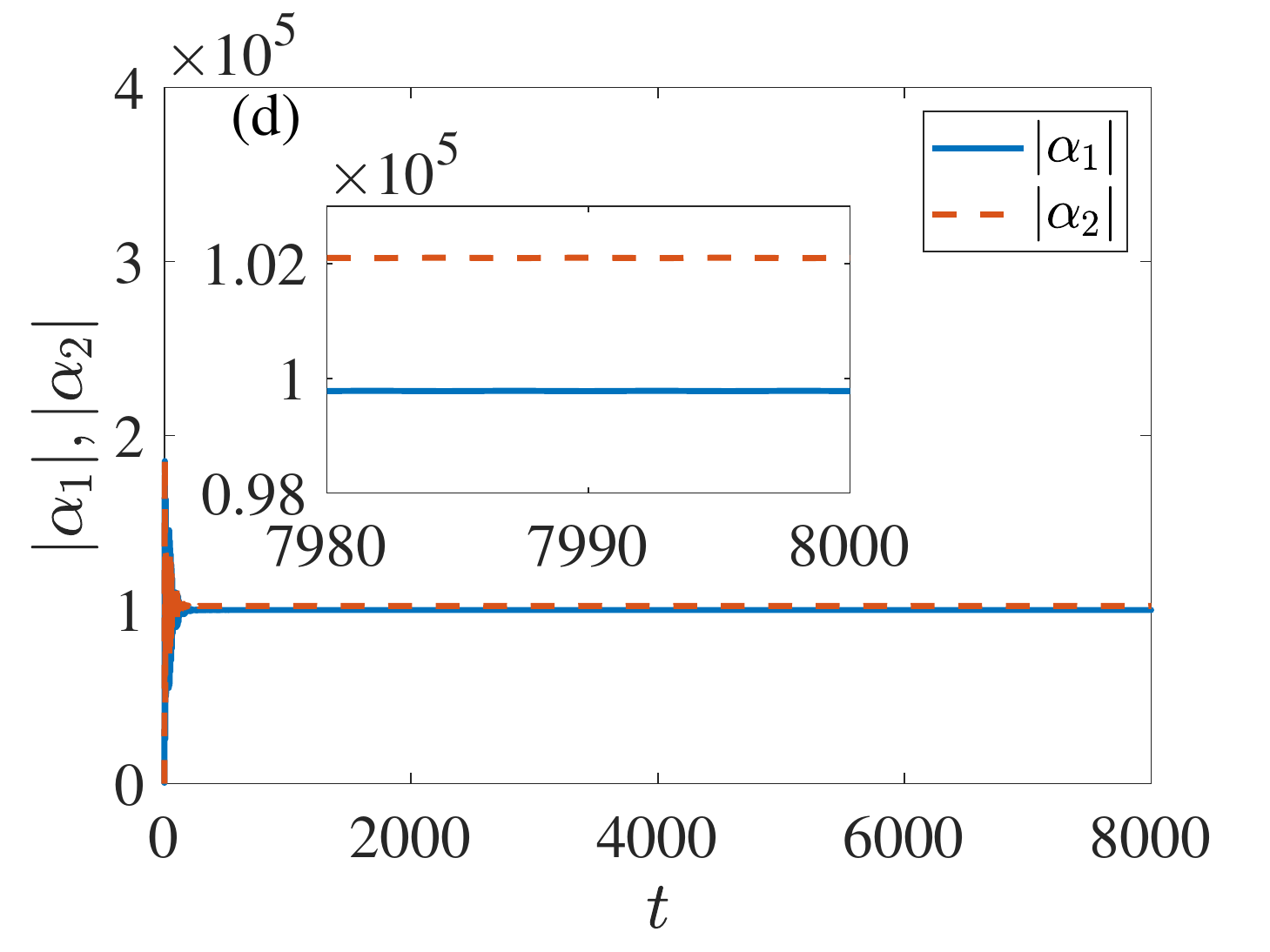}}
	\subfigure{\includegraphics[width=0.32\linewidth]{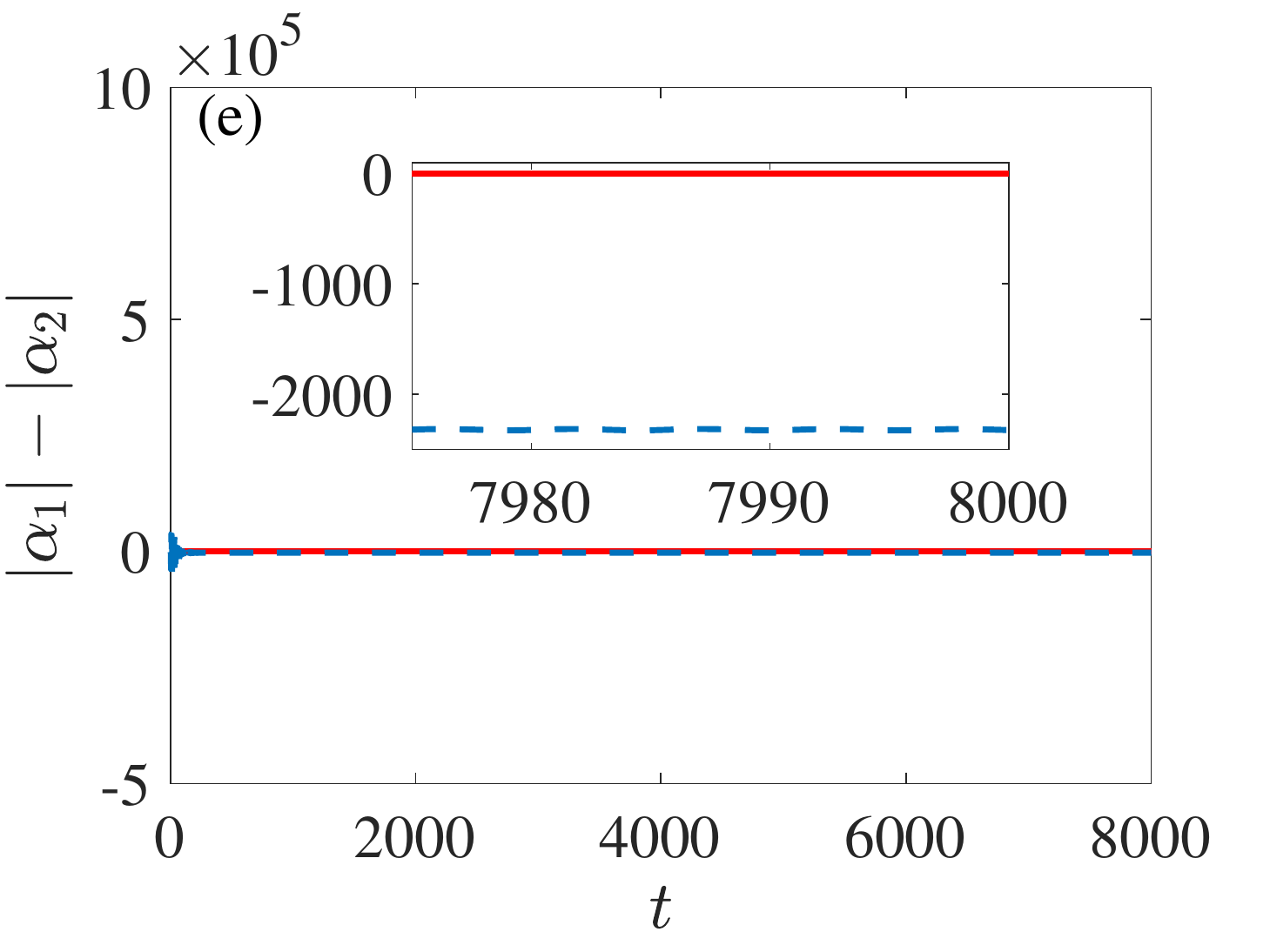}}
	\subfigure{\includegraphics[width=0.32\linewidth]{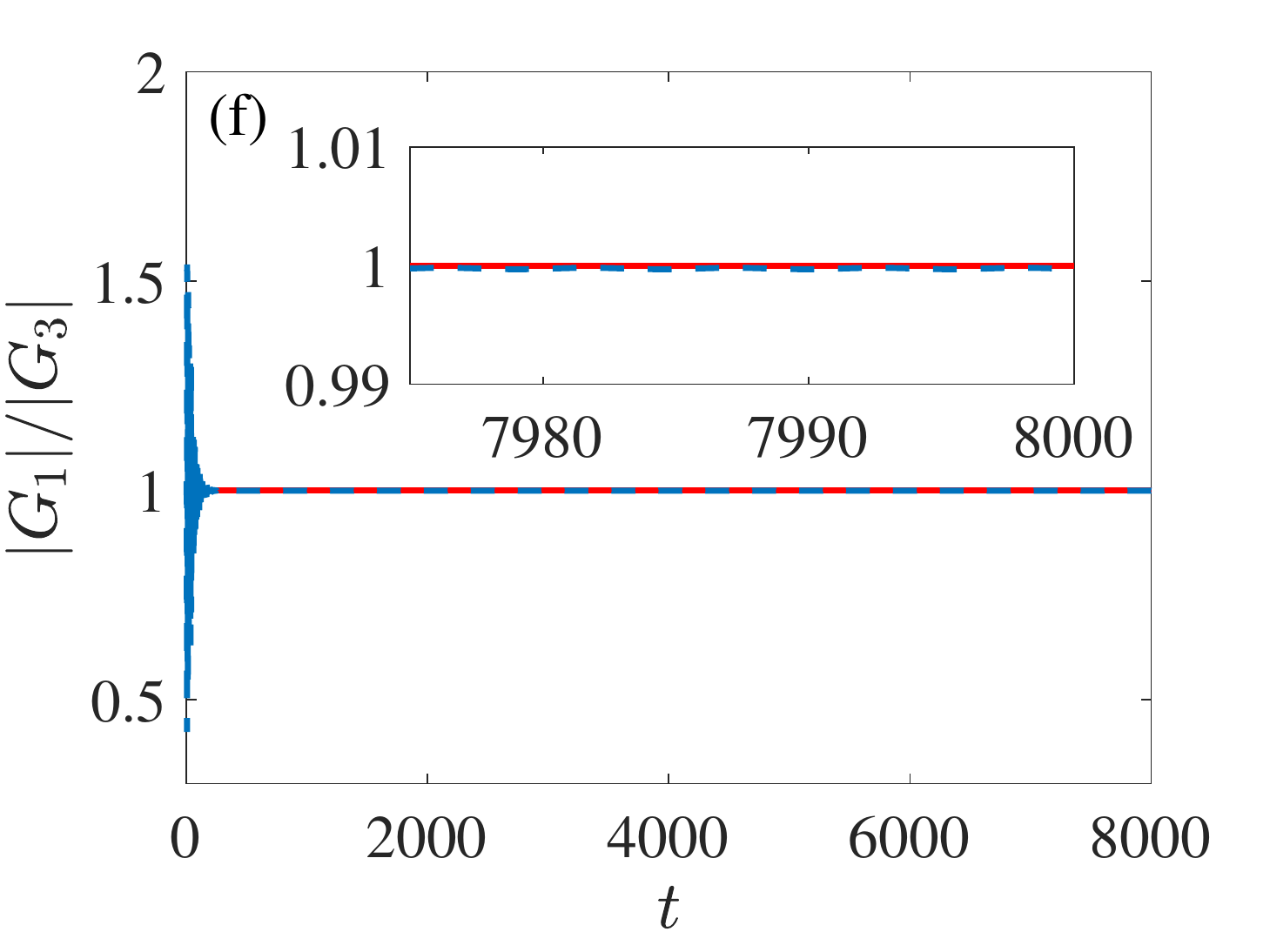}}
	\caption{The steady-state cavity fields versus the time $t$. (a) The distributions of the steady-state cavity fields $\alpha_{1}$ and $\alpha_{2}$. (b) The difference between the steady-state cavity field $\alpha_{1}$ and the steady-state cavity field $\alpha_{2}$. (c) The ratio between the steady-state cavity field $\alpha_{1}$ and the steady-state cavity field $\alpha_{2}$. In (a), (b), and (c), the parameters take $\omega_{b,1}=\omega_{b,2}=\omega_{b}$, $\Delta_{a,1}=\Delta_{a,2}=\omega_{b}$, $g_{1}=g_{2}=1\times10^{-6}\omega_{b}$, $\Omega_{1}=\Omega_{2}=1\times10^{5}\omega_{b}$, $\kappa_{1}=\kappa_{2}=0.1\omega_{b}$, and $\gamma_{1}=\gamma_{2}=1\times10^{-5}\omega_{b}$. (d) The distributions of the steady-state cavity fields $\alpha_{1}$ and $\alpha_{2}$. (e) The difference between the steady-state cavity field $\alpha_{1}$ and the steady-state cavity field $\alpha_{2}$. (f) The ratio between the effective optomechanical coupling $|G_{1}|$ and the effective optomechanical coupling $|G_{3}|$. In (d), (e), and (f),  $g_{1}=1.023\times10^{-6}\omega_{b}$, $g_{2}=1.0\times10^{-6}\omega_{b}$, and other parameters are the same as the parameters in (a), (b), and (c). We set $\omega_{b}=1$ as the energy unit.}\label{fig2}
\end{figure*}  
We consider a one dimensional (1D) small optomechanical lattice consisting of two optical cavity fields and two resonators, in which the two optical cavity fields are driven by two external lasers with the driving amplitudes $\Omega_{1}$ and $\Omega_{2}$, as shown in Fig.~\ref{fig1}. Then, the system can be described by the Hamiltonian 
\begin{eqnarray}\label{e01}
H&=&\sum_{j=1,2}\left[\omega_{a,j}a_{j}^{\dag}a_{j}+\omega_{b,j}b_{j}^{\dag}b_{j}\right]\cr\cr
&&+\sum_{j=1,2}\left[\Omega_{j}a_{j}^{\dag}e^{-i\omega_{d,j}t}+\Omega_{j}^{\ast}a_{j}e^{i\omega_{d,j}t}\right]\cr\cr
&&-\left[g_{1}(a_{1}^{\dag}a_{1}-a_{2}^{\dag}a_{2})(b_{1}^{\dag}+b_{1})+g_{2}a_{2}^{\dag}a_{2}(b_{2}^{\dag}+b_{2})\right],\cr
&&
\end{eqnarray}
where $a_{n}^{\dag}$ ($a_{n}$) and $b_{n}^{\dag}$ ($b_{n}$) are the creation (annihilation) operators of the cavity fields and resonators. The first summation of the above Hamiltonian is the free energy of the cavity fields and the resonators with the frequency $\omega_{a,j}$ and $\omega_{b,j}$. The second summation represents the driving term of the external lasers with the driving frequency $\omega_{d,j}$. And the last term is the coupling between the cavity fields and the resonators with the single-photon optomechanical couplings $g_{1}$ and $g_{2}$. After performing a rotating transformation defined by the external driving frequency, the Hamiltonian becomes
\begin{eqnarray}\label{e02}
H_{1}&=&\sum_{j=1,2}\left[\Delta_{a,j}a_{j}^{\dag}a_{j}+\omega_{b,j}b_{j}^{\dag}b_{j}+\Omega_{j}a_{j}^{\dag}+\Omega_{j}^{\ast}a_{j}\right]\cr\cr
&&-\left[g_{1}(a_{1}^{\dag}a_{1}-a_{2}^{\dag}a_{2})(b_{1}^{\dag}+b_{1})+g_{2}a_{2}^{\dag}a_{2}(b_{2}^{\dag}+b_{2})\right],\cr
&&
\end{eqnarray}   
where $\Delta_{a,j}=\omega_{a,j}-\omega_{d,j}$ is the detuning between the cavity fields and the external lasers. To further investigate the steady-state dynamics of the small optomechanical lattice, we implement the standard linearization process via rewriting the operators as $a_{j}=\left \langle a_{j} \right \rangle+\delta a_{j}=\alpha_{j}+\delta a_{j}$ ($b_{j}=\left \langle b_{j} \right \rangle+\delta b_{j}=\beta_{j}+\delta b_{j}$). After dropping the notation ``$\delta$'' for all the fluctuation operators $\delta a_{j}$ ($\delta b_{j}$), the Hamiltonian can be rewritten as
\begin{eqnarray}\label{e03}
H_{1}&=&\sum_{j=1,2}\left[\Delta_{a,j}^{'}a_{j}^{\dag}a_{j}+\omega_{b,j}b_{j}^{\dag}b_{j}\right]\cr\cr
&&-\left[g_{1}(\alpha_{1}a_{1}^{\dag}+\alpha_{1}^{\ast}a_{1}-\alpha_{2}a_{2}^{\dag}-\alpha_{2}^{\ast}a_{2})(b_{1}^{\dag}+b_{1})\right.\cr\cr
&&\left.+g_{2}(\alpha_{2}a_{2}^{\dag}+\alpha_{2}^{\ast}a_{2})(b_{2}^{\dag}+b_{2})\right],\cr
&&
\end{eqnarray}  
where $\Delta_{a,1}^{'}=\Delta_{a,1}-g_{1}(\beta_{1}+\beta_{1}^{\ast})$ and $\Delta_{a,2}^{'}=\Delta_{a,2}+g_{1}(\beta_{1}+\beta_{1}^{\ast})-g_{2}(\beta_{2}+\beta_{2}^{\ast})$ are the effective cavity field detuning originating from the optomechanical coupling.
Under the red-detuning condition, the counter rotating wave terms can be removed effectively, and the final effective Hamiltonian can be written as  
\begin{eqnarray}\label{e04}
H_{eff}&=&\left[-G_{1}a_{1}^{\dag}b_{1}+G_{2}a_{2}^{\dag}b_{1}-G_{3}a_{2}^{\dag}b_{2}\right]+\mathrm{H.c.},
\end{eqnarray}  
where $G_{1}=g_{1}\alpha_{1}$, $G_{2}=g_{1}\alpha_{2}$, and $G_{3}=g_{2}\alpha_{2}$ are the effective optomechanical couplings with $\alpha_{1}$, $\alpha_{2}$, and $\alpha_{3}$ being the final steady-state cavity fields. The above Hamiltonian only possesses the interactions between the two adjacent cavity field and the resonator, which is equivalent to the tight-binding Hamiltonian in form. It means that the present small optomechanical lattice can be mapped into a topological tight-binding model depending on the different values of the effective optomechanical couplings. 

\section{\label{sec.3} Topological phase and phase transition based on the small optomechanical lattice}
As mentioned above, the different forms of the effective optomechanical couplings determine the topology of the optomechanical lattice. Meanwhile, the effective optomechanical couplings are closely related to the final steady-state cavity fields, implying that the final steady-state cavity fields need to be analyzed. The final steady-state cavity fields can be determined by the following steady-state equations, with
\begin{eqnarray}\label{e05}
\dot{\alpha_{1}}&=&-i[\Delta_{a,1}-g_{1}(\beta_{1}+\beta_{1}^{\ast})]\alpha_{1}-i\Omega_{1}-\frac{\kappa_{1}}{2}\alpha_{1},\cr\cr
\dot{\beta_{1}}&=&-i(\omega_{b,1}\beta_{1}-g_{1}|\alpha_{1}|^{2}+g_{1}|\alpha_{2}|^{2})-\frac{\gamma_{1}}{2}\beta_{1},\cr\cr
\dot{\alpha_{2}}&=&-i[\Delta_{a,2}+g_{1}(\beta_{1}+\beta_{1}^{\ast})-g_{2}(\beta_{2}+\beta_{2}^{\ast})]\alpha_{2}\cr\cr
&&-i\Omega_{2}-\frac{\kappa_{2}}{2}\alpha_{2},\cr\cr
\dot{\beta_{2}}&=&-i(\omega_{b,2}\beta_{2}-g_{2}|\alpha_{2}|^{2})-\frac{\gamma_{2}}{2}\beta_{2},
\end{eqnarray}  
where $\kappa_{j}$ and $\gamma_{j}$ (with $j=1,2$) are the decay of the cavity fields and the damping of the resonators, respectively. The above steady-state equations indicate that, after a long time of evolution, the system may enter into the steady state due to the existence of the decay of the cavity fields. In the following, we focus on the effects of the final steady-state cavity fields on the effective optomechanical coupling and the topology of the small optomechanical lattice. 
\begin{figure}
	\centering
	\subfigure{\includegraphics[width=0.49\linewidth]{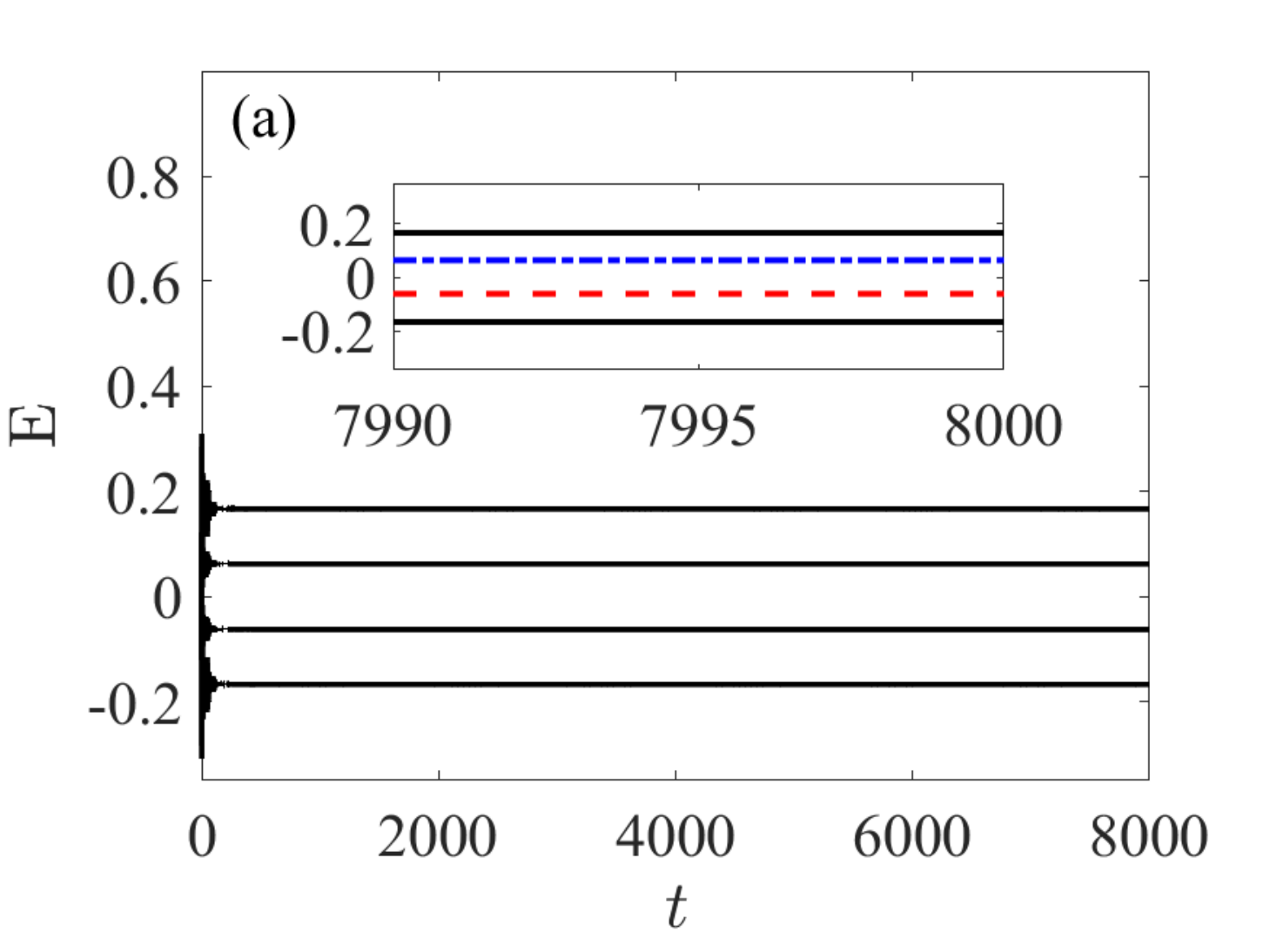}}
	\subfigure{\includegraphics[width=0.49\linewidth]{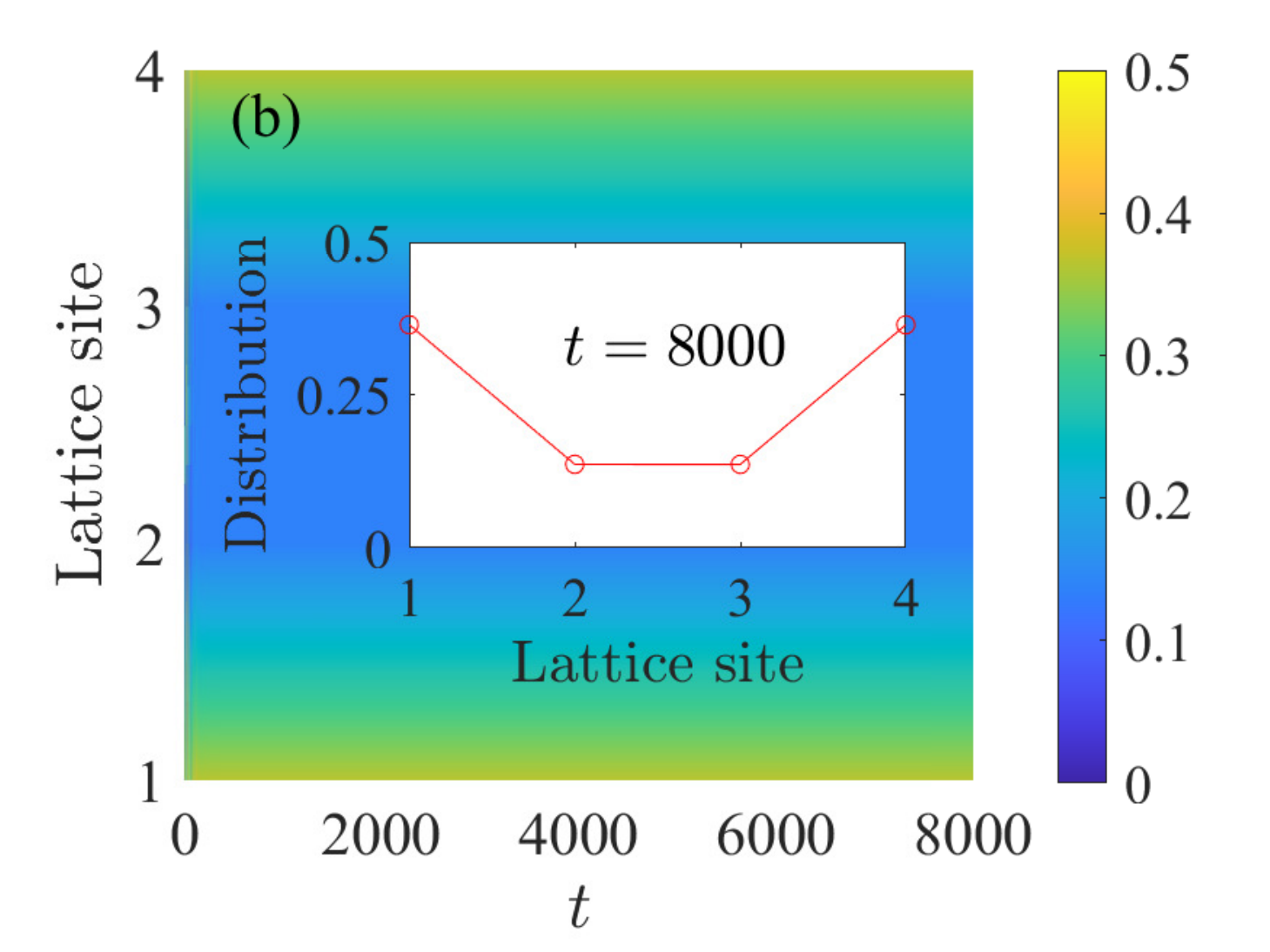}}		
	\caption{The energy spectrum and the distribution of gap state. (a) The energy spectrum of the SSH model when the intra-cell and inter-cell coupling satisfy $|G_{1}|=|G_{3}|<|G_{2}|$. (b) The distribution of the blue gap state in (a), in which the gap state has the maximal distributions at the two ends cavity fields. The parameters are the same as the parameters in Figs.~\ref{fig2}(d)-\ref{fig2}(f). We set $\omega_{b}=1$ as the energy unit.}\label{fig3}
\end{figure} 
\begin{figure}
	\centering
	\subfigure{\includegraphics[width=0.49\linewidth]{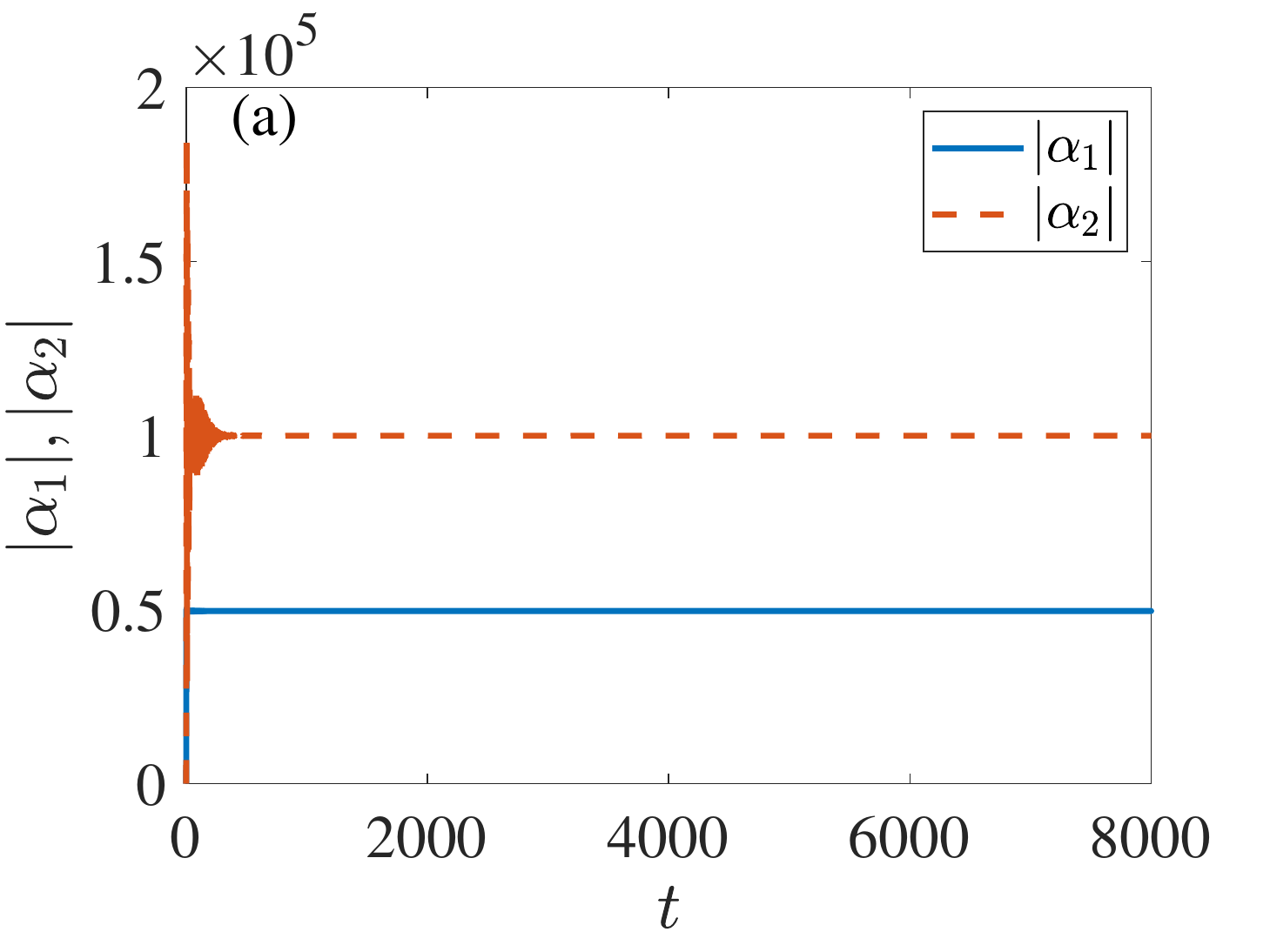}}
	\subfigure{\includegraphics[width=0.49\linewidth]{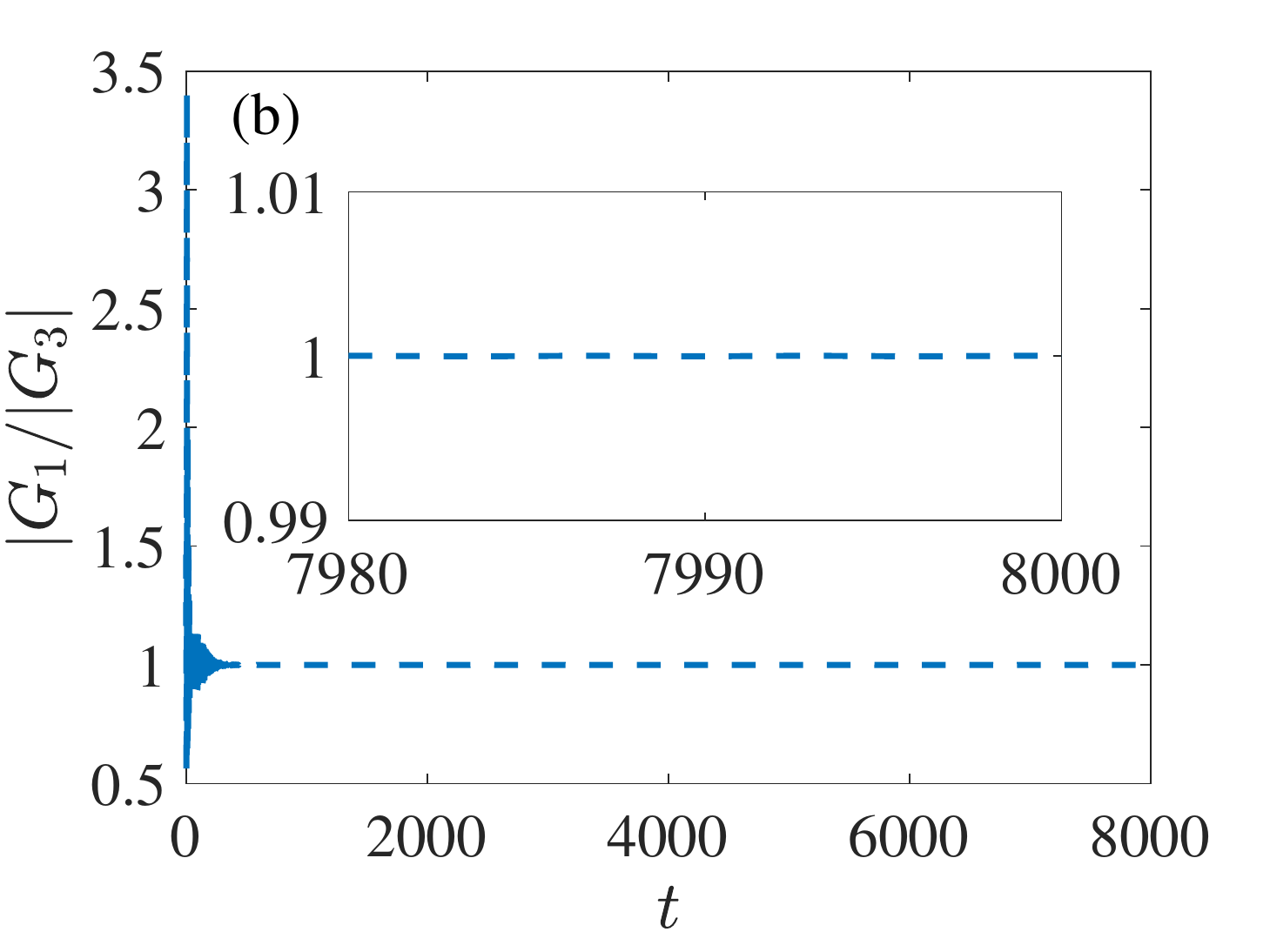}}	
	
	\subfigure{\includegraphics[width=0.49\linewidth]{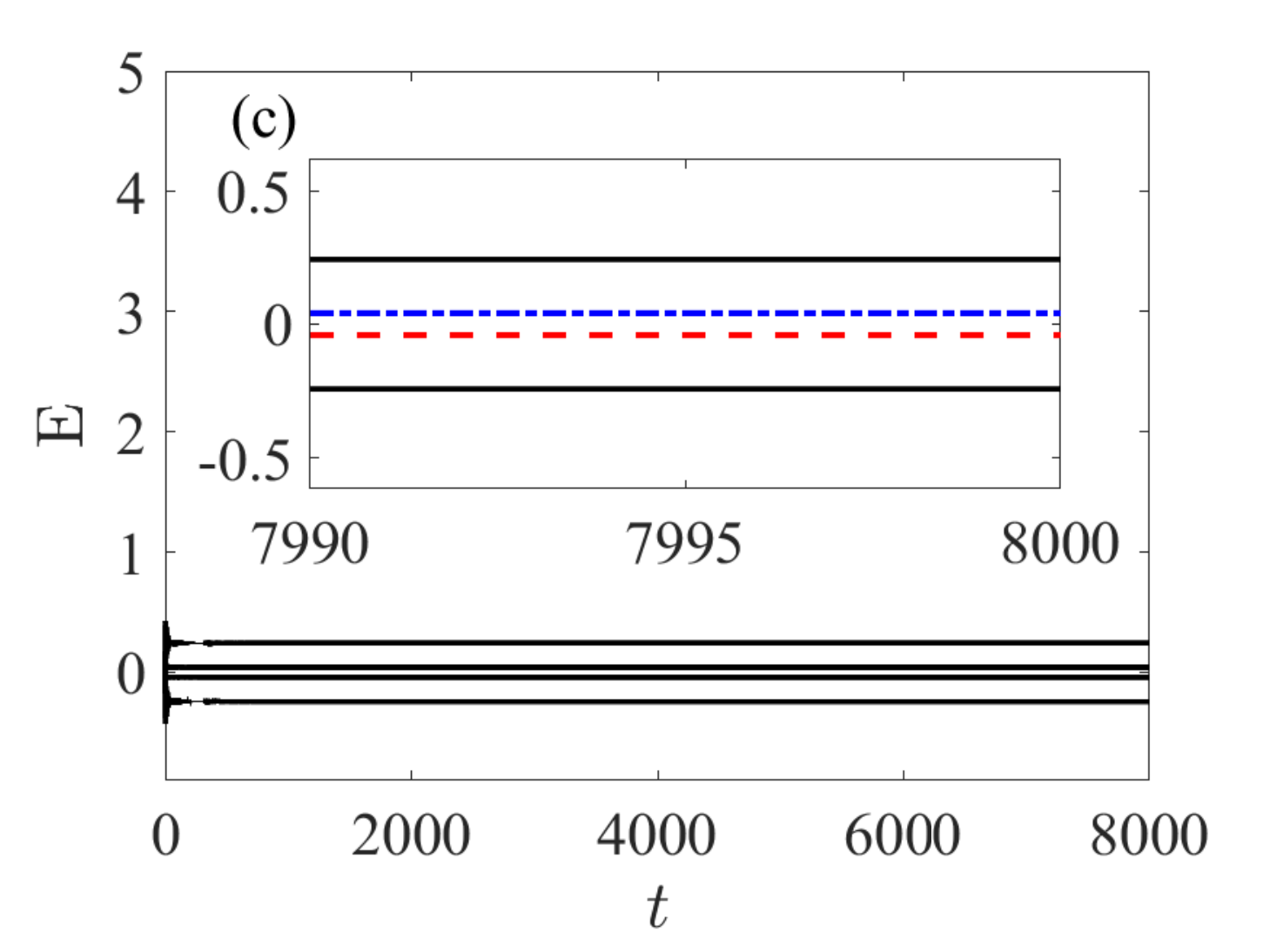}}
	\subfigure{\includegraphics[width=0.49\linewidth]{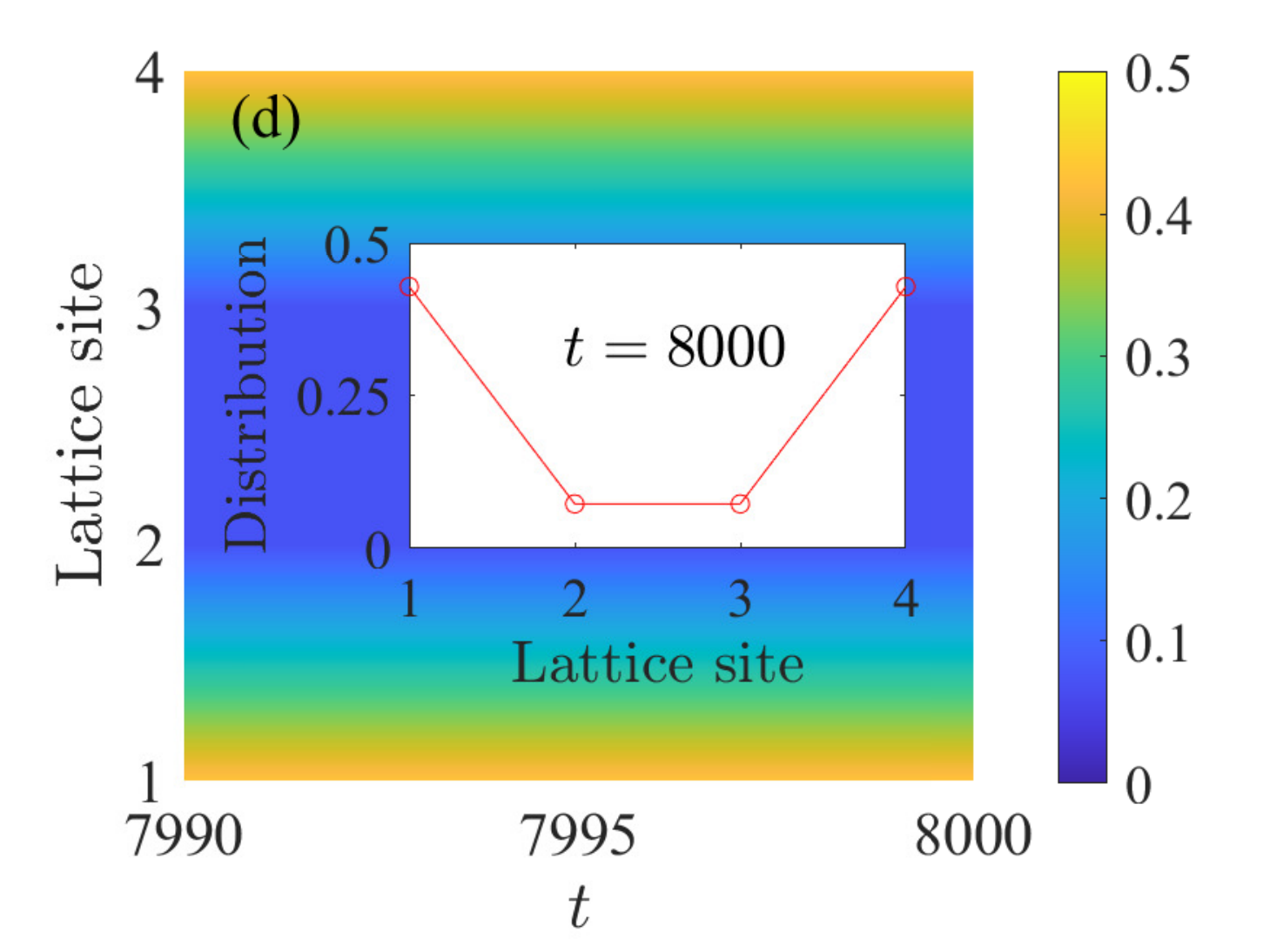}}	
	\caption{The steady state dynamics of the small optomechanical lattice when $\kappa_{1}\gg\kappa_{2}$. (a) The two final steady cavity fields versus the time $t$. (b) The ratio between the effective optomechanical coupling $|G_{1}|$ and $|G_{3}|$. (c) The energy spectrum of the small optomechanical lattice. The two colored lines represent the two gap states. (d) The distribution of the blue gap state in (c), in which the gap state has the maximal distributions at the two ends cavity fields. The maximal distributions at the two ends cavity fields are larger than the distributions in Fig.~\ref{fig3}. The parameters take $g_{1}=2.015\times10^{-6}\omega_{b}$, $g_{2}=1.0\times10^{-6}\omega_{b}$, $\kappa_{1}=3.5\omega_{b}$, and $\kappa_{2}=0.1\omega_{b}$. Other parameters are the same as parameters in Fig.~\ref{fig3}. We set $\omega_{b}=1$ as the energy unit.}\label{fig4}
\end{figure} 
\begin{figure}
	\centering
	\subfigure{\includegraphics[width=0.49\linewidth]{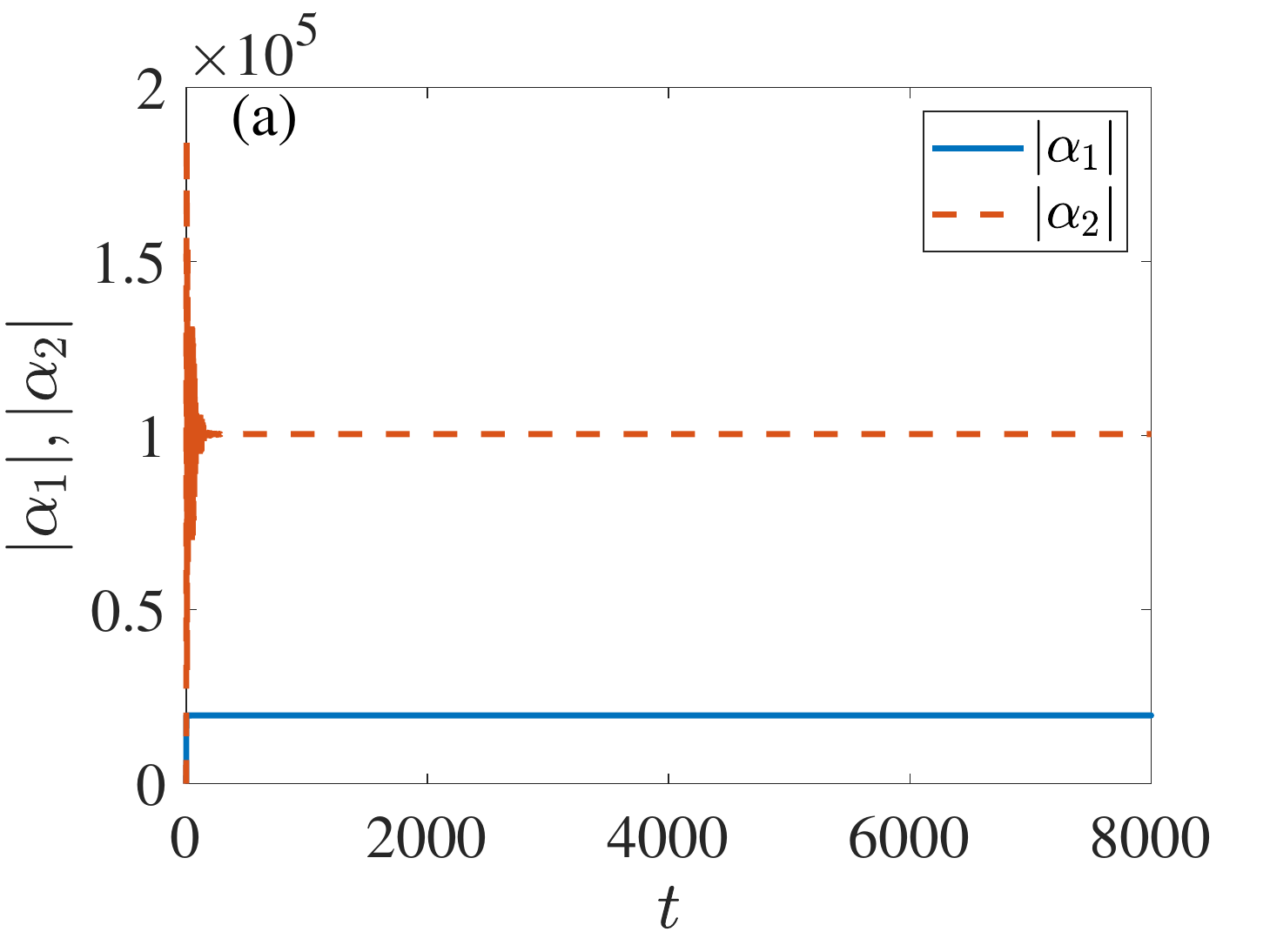}}
	\subfigure{\includegraphics[width=0.49\linewidth]{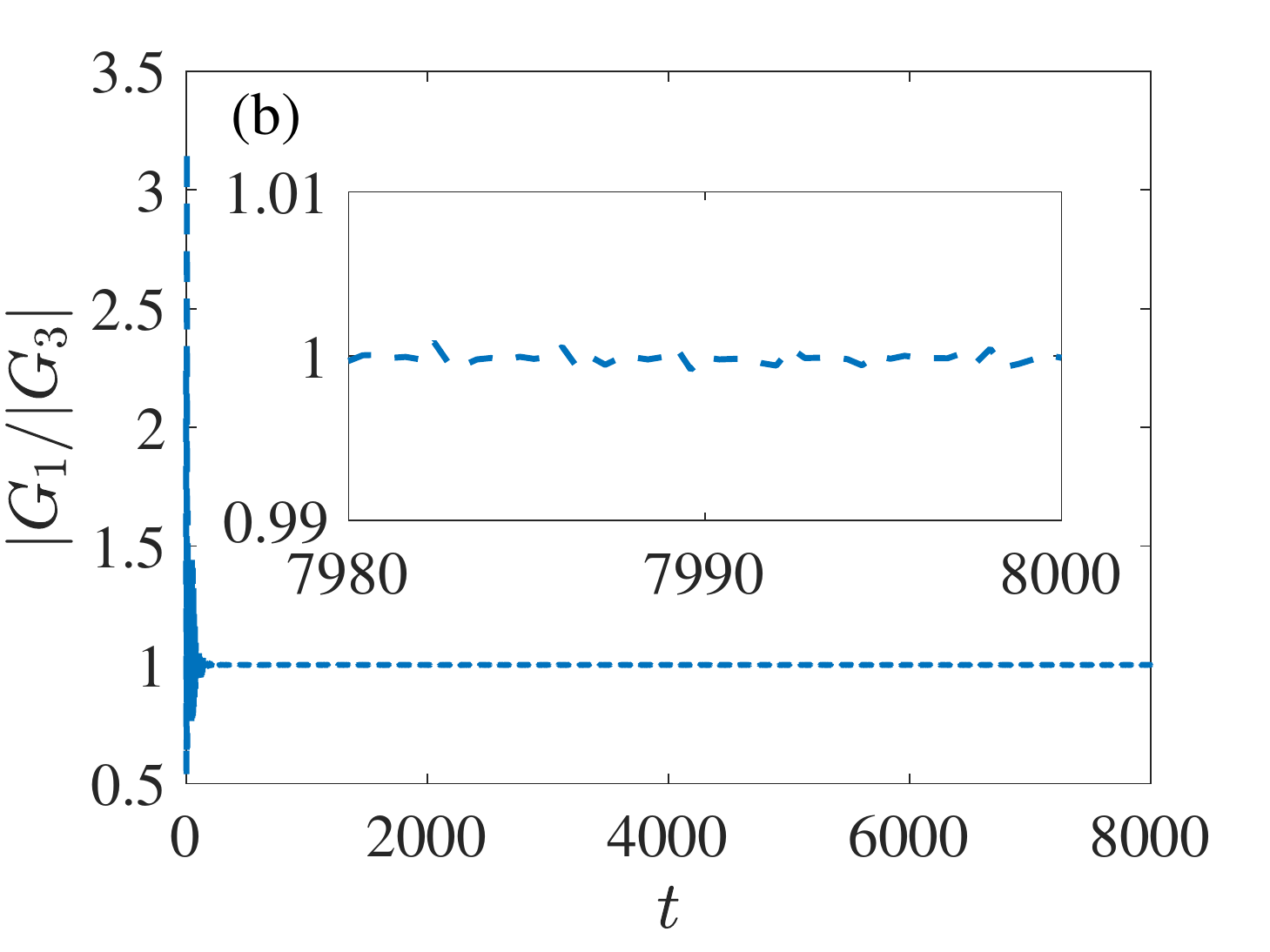}}	
	
	\subfigure{\includegraphics[width=0.49\linewidth]{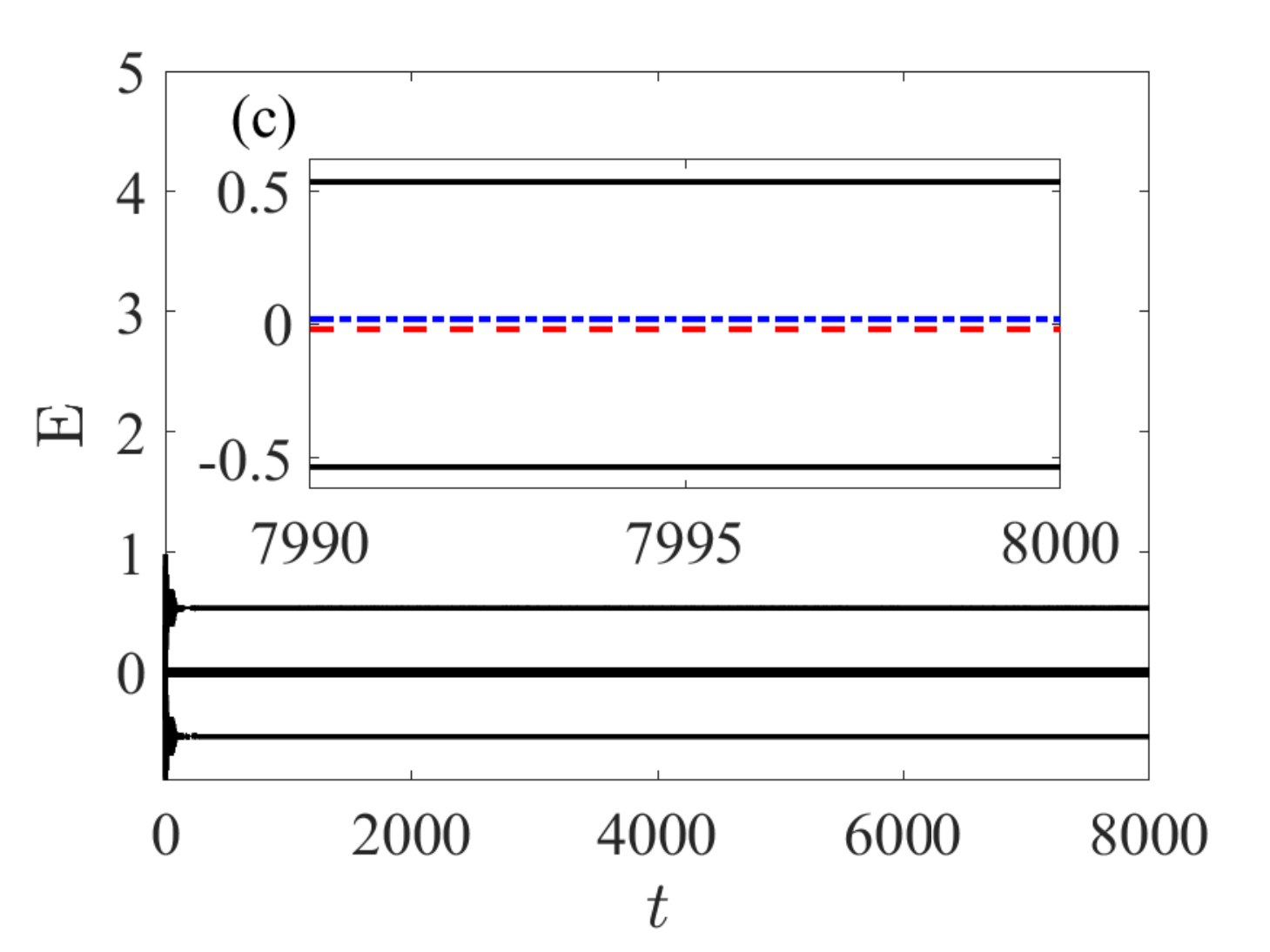}}
	\subfigure{\includegraphics[width=0.49\linewidth]{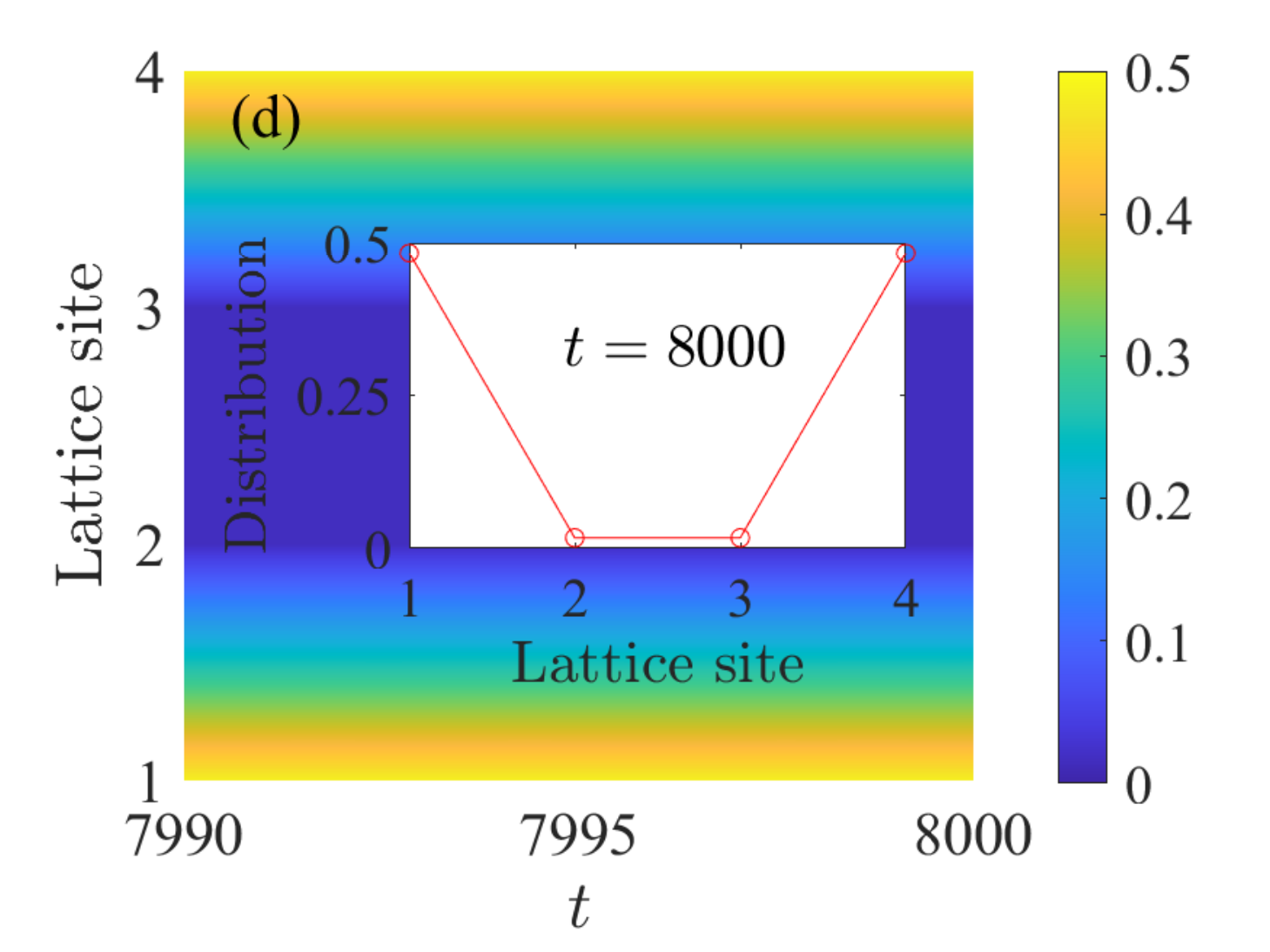}}	
	\caption{The steady state dynamics of the small optomechanical lattice under the large $\kappa_{1}$ limit. (a) The two final steady cavity fields versus the time $t$. (b) The ratio between the effective optomechanical coupling $|G_{1}|$ and $|G_{3}|$. (c) The energy spectrum of the small optomechanical lattice. The energy spectrum approximately has two degenerate zero energy modes. (d) The distribution of the blue gap state in (c), in which the gap state only has the maximal distributions at the two ends cavity fields. The parameters take $g_{1}=5.12\times10^{-6}\omega_{b}$, $g_{2}=1.0\times10^{-6}\omega_{b}$, $\kappa_{1}=10\omega_{b}$, and $\kappa_{2}=0.1\omega_{b}$. Other parameters are the same as parameters in Fig.~\ref{fig3}. We set $\omega_{b}=1$ as the energy unit.}\label{fig5}
\end{figure} 
\begin{figure}
	\centering
	\subfigure{\includegraphics[width=0.49\linewidth]{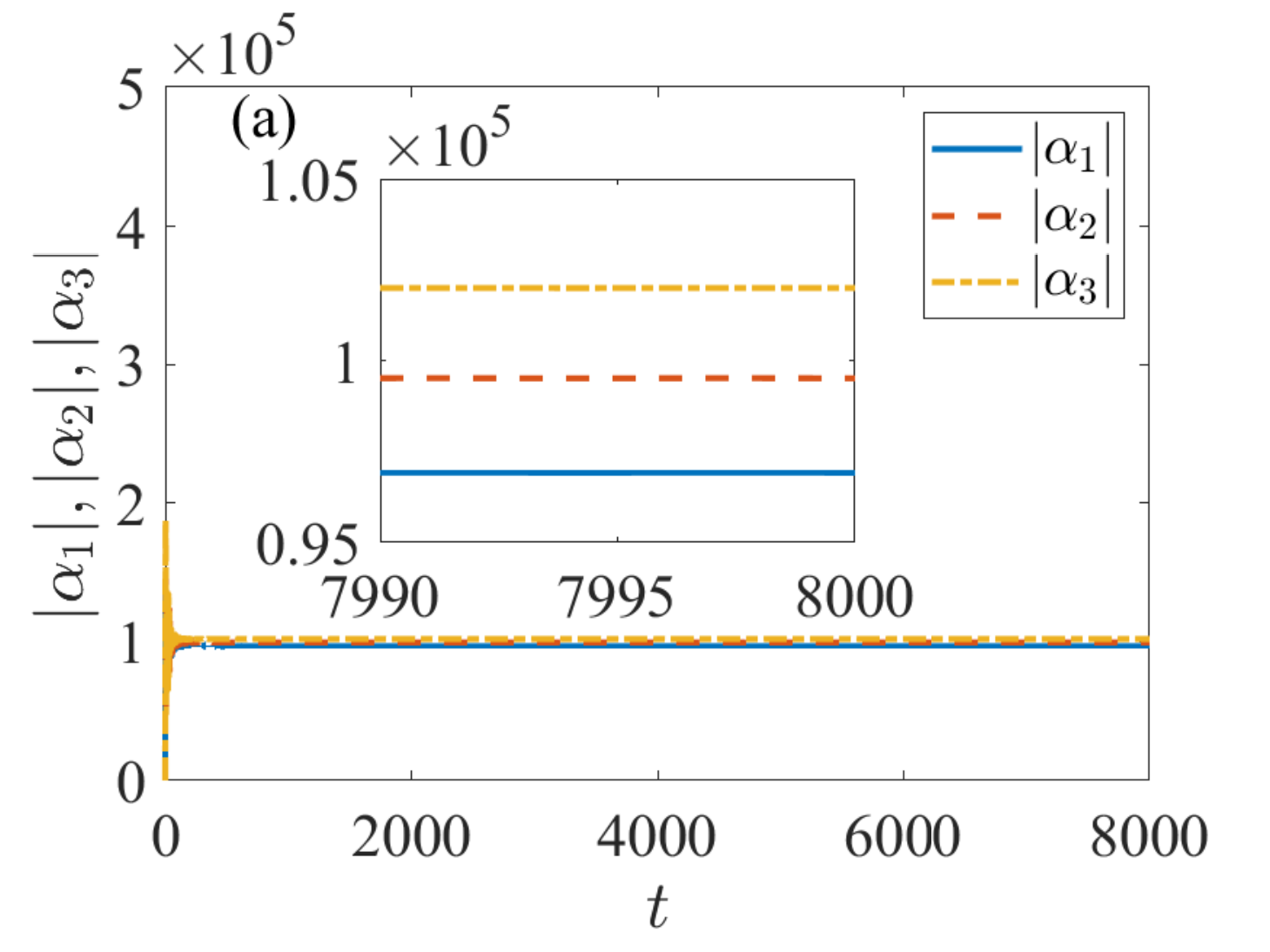}}
	\subfigure{\includegraphics[width=0.49\linewidth]{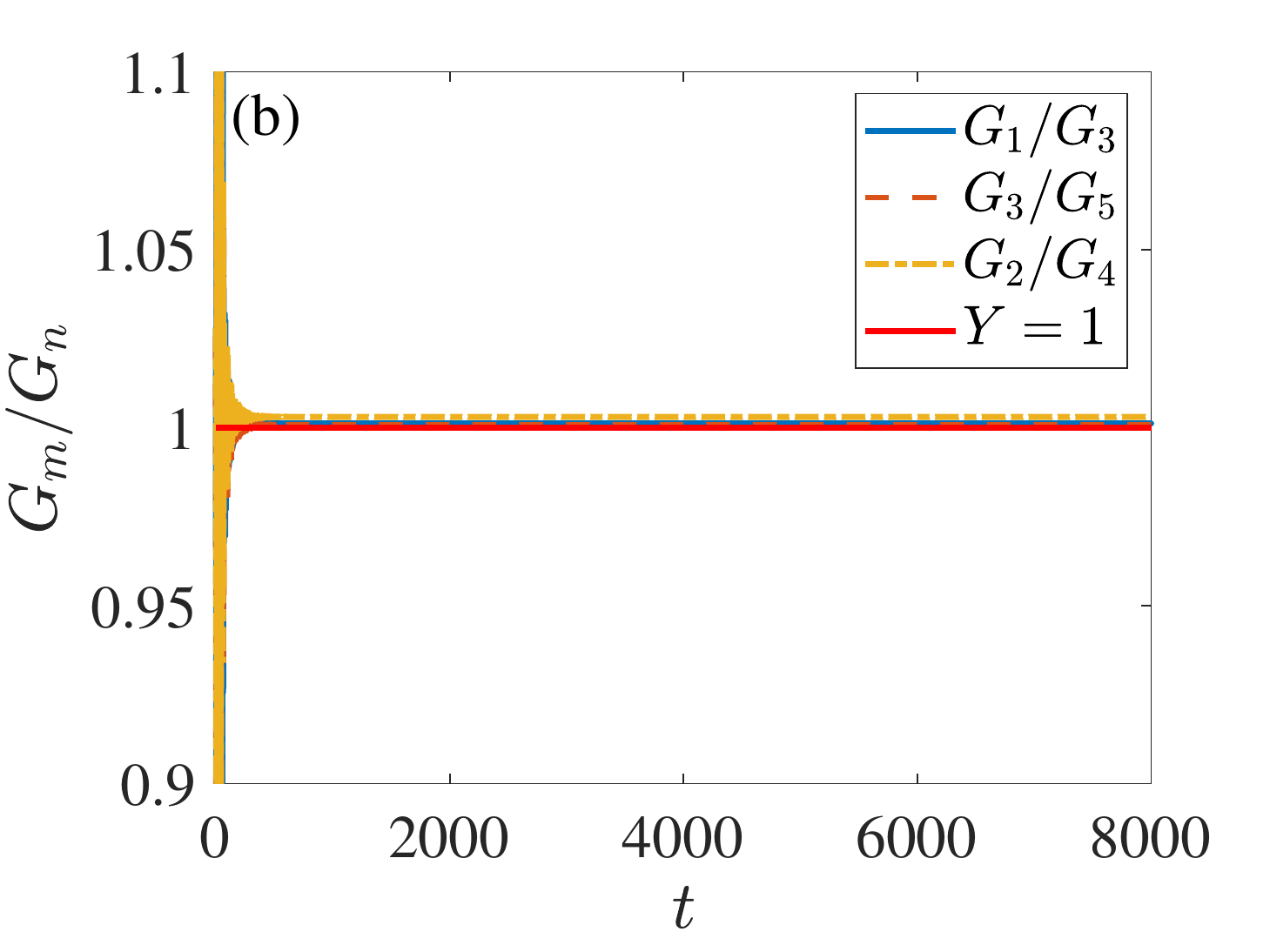}}	
	\caption{The steady state dynamics of the small optomechanical lattice with three unit cells. (a) The three final steady cavity fields versus the time $t$. (b) The ratio between the two effective optomechanical coupling. We have $|-G_{1}|=|G_{3}|=|-G_{5}|$ and $|G_{2}|=|G_{4}|$ approximately, which ensures the coupling configuration of the SSH model. The parameters take $g_{1}=1.028\times10^{-6}\omega_{b}$, $g_{2}=1.0\times10^{-6}\omega_{b}$, $g_{3}=0.975\times10^{-6}\omega_{b}$, $\kappa_{1}=0.5\omega_{b}$, $\kappa_{2}=0.2\omega_{b}$, and $\kappa_{3}=0.1\omega_{b}$. Other parameters are the same as parameters in Fig.~\ref{fig3}. We set $\omega_{b}=1$ as the energy unit.}\label{fig6}
\end{figure}

\subsection{\label{sec.A} Mapping of topological Su-Schrieffer-Heeger model}
We first consider a parameter regime, in which the key parameters satisfy $g_{1}=g_{2}$ and $\kappa_{1}=\kappa_{2}$. We find that the final steady-state cavity fields $\alpha_{1}$ and $\alpha_{2}$ are indeed stabilized at fixed values, as shown in Fig.~\ref{fig2}(a). Especially, under the steady state regime, the final steady-state cavity fields satisfy $|\alpha_{1}|<|\alpha_{2}|$ at any moment, as shown in Fig.~\ref{fig2}(b). The phenomenon is spontaneously induced by the physical structure of the small optomechanical lattice, namely, the second cavity field couples two resonators at the same time comparing with the first cavity field. Thus, the effective optomechanical couplings satisfy $|-g_{1}\alpha_{1}|<|g_{1}\alpha_{2}|$ ($|G_{1}|<|G_{2}|$), which means that the intra-cell hopping is weaker than the inter-cell hopping if we treat the cavity field $a_{1}$ ($a_{2}$) and resonator $b_{1}$ ($b_{2}$) as one unit cell. It indicates that the effective tight-binding Hamiltonian in Eq.~(\ref{e04}) can be potentially mapped into a topological SSH model. However, the premise is that the third effective optomechanical coupling $G_{3}$ must satisfy $|G_{3}|=|G_{1}|$, which is obviously impossible since $\frac{|G_{1}|}{|G_{3}|}=\frac{|\alpha_{1}|}{|\alpha_{2}|}<1$, as shown in Fig.~\ref{fig2}(c). Thus, we need further to modulate the optomechanical coupling $g_{1}$ and $g_{2}$ to ensure the condition of $|G_{3}|=|G_{1}|$, such as $g_{1}=1.203g_{2}$. The final steady-state cavity fields, under the new parameter regime, are still stabilized at fixed values and satisfy $|\alpha_{1}|<|\alpha_{2}|$ ($|G_{1}|<|G_{2}|$), as shown in Figs.~\ref{fig2}(d) and~\ref{fig2}(e). Especially, as shown in Fig.~\ref{fig2}(f), the third effective coupling now satisfies $|G_{3}|=|G_{1}|$. In this way, the final effective tight-binding Hamiltonian in Eq.~(\ref{e04}) becomes
\begin{eqnarray}\label{e06}
H&=&\left[-G_{1}a_{1}^{\dag}b_{1}+G_{2}a_{2}^{\dag}b_{1}-G_{1}a_{2}^{\dag}b_{2}\right]+\mathrm{H.c.}.
\end{eqnarray}      
Obviously, the above Hamiltonian is equivalent to a tight-binding SSH-type Hamiltonian only possessing two unit cells with intra-cell coupling $-G_{1}$ and inter-cell coupling $G_{2}$. As shown in Figs.~\ref{fig2}(d) and~\ref{fig2}(e), the numerical results indicate that, under the steady state, the intra-cell coupling $-G_{1}$ and inter-cell coupling $G_{2}$ always satisfy $|-G_{1}|<|G_{2}|$, which means that the small optomechanical lattice has two edge modes locating at the first cavity field and the last resonator respectively. To further clarify it, we plot the energy spectrum of the SSH model and the distribution of the gap state, as shown in Fig.~\ref{fig3}. The numerical results reveal that, the energy spectrum of the present tight-binding SSH model indeed has two nonzero gap states locating into the gap, as shown in Fig.~\ref{fig3}(a). And the gap state of the SSH model has the maximal distributions at the two ends cavity fields, as shown in Fig.~\ref{fig3}(b). Note that the gap state also has certain distributions at bulk sites, implying the insufficiently topological edge localizations. The reason is that, as shown in Figs.~\ref{fig2}(d) and~\ref{fig2}(e), although the effective optomechanical couplings satisfy $|G_{1}|<|G_{2}|$, the difference between two effective optomechanical couplings is relatively small. As well known, in the usual SSH model, the larger difference between the inter-cell and intra-cell couplings corresponds to the stronger localized effects of the edge states. Thus, the relatively small difference between the inter-cell and intra-cell couplings leads that the localized effects of the edge states are weakened. Besides, the present optomechanical lattice only has four lattice sites, which indicates that the finite-size effect~\cite{obana2019topological,liu2018topological,vos1996su} further aggravates the weakening of the localized effects.

\subsection{\label{sec.B} Strengthened topological effects induced via two different ways}
As mentioned above, the condition of $|-G_{1}|=|-G_{3}|<|G_{2}|$ leads that the two gap states are separated by a finite gap and the localized effects of the gap states are weakened. Actually, we have two ways to avoid the weakened topologically localized effects of the gap states, in which one way is that we take the condition of $|-G_{1}|=|-G_{3}|\ll |G_{2}|$ to ensure that the inter-cell and intra-cell couplings have large enough difference, while, another way is that we increase the size of the system directly. 

The condition $|-G_{1}|=|-G_{3}|\ll |G_{2}|$ actually implies that the parameters should satisfy $|\alpha_{1}|\ll |\alpha_{2}|$ and $g_{1}\gg g_{2}$. In the optomechanical system, the photons will eventually decay into the vacuum accompanying with a smaller final steady cavity field when the cavity field takes a large decay rate. Thus, the condition $|\alpha_{1}|\ll |\alpha_{2}|$ can be realized via choosing the decay of the cavity fields to satisfy $\kappa_{1}\gg \kappa_{2}$. For example, when $\kappa_{1}=3.5\omega_{b}$ and $\kappa_{2}=0.1\omega_{b}$, we find that the two final steady cavity fields indeed satisfy $\alpha_{1}\ll \alpha_{2}$, as shown in Fig.~\ref{fig4}(a). In this way, we approximately have $|-g_{1}\alpha_{1}|\ll |g_{1}\alpha_{2}|$, namely, $|-G_{1}|\ll |G_{2}|$. Note that we still need the parameters to satisfy $|-G_{1}|=|G_{3}|$ to realize the mapping of the SSH model, which can be easily realized via choosing the optomechanical coupling $g_{1}$ and $g_{2}$ as $g_{1}=2.015\times10^{-6}\omega_{b}$ and $g_{2}=1\times10^{-6}\omega_{b}$. The numerical results of $\frac{|G_{1}|}{|G_{3}|}$ versus the time $t$ are plotted in Fig.~\ref{fig4}(b). Obviously, under the present parameter regime, the condition of $|G_{1}|=|G_{3}|$ can be satisfied, implying that the small optomechanical lattice can be mapped into a topologically nontrivial SSH model. To further verify it, we plot the energy spectrum and the distribution of the gap state, as shown in Figs.~\ref{fig4}(c) and~\ref{fig4}(d). The numerical results reveal that, as shown in Fig.~\ref{fig4}(c), the larger decay $\kappa_{1}=3.5\omega_{b}$ leads that the two originally separated gap states move toward each other. Meanwhile, the localizations of gap state at two ends sites are strengthened while the distributions of gap state at bulk sites are weakened, as shown in Fig.~\ref{fig4}(d).     

Similarly, we can further increase the decay of the first cavity to further strengthen the topological effect of the small optomechanical lattice. The numerical results, under the large $\kappa_{1}$ limit with $\kappa_{1}=10\omega_{b}$, are shown in Fig.~\ref{fig5}. We find that the final steady-state cavity field of the first cavity indeed further decreases, inducing the larger difference between the intra-cell coupling $-G_{1}$ and inter-cell coupling $G_{2}$, as shown in Fig.~\ref{fig5}(a). At the same time, we can also ensure the condition of $|G_{1}|=|G_{3}|$ to realize the mapping of the SSH model, as shown in Fig.~\ref{fig5}(b). The energy spectrum and the distribution of the gap state are shown in Figs.~\ref{fig5}(c) and~\ref{fig5}(d). The numerical results reveal that the large cavity decay limit $\kappa_{1}=10\omega_{b}$ leads that the two originally separated gap states become degenerate approximately and the gap state is localized at two ends completely. These phenomena imply that the present small optomechanical lattice can be mapped into a topologically nontrivial SSH phase.    

In addition to the condition of $|-G_{1}|=|-G_{3}|\ll|G_{2}|$, another way to suppress the effects of the size effect on the gap states is that, we increase the size of the optomechanical lattice directly depending on the scalability of the single optomechanical system. In this way, for a large enough optomechanical lattice with $N$ unit cells, we should ensure $G_{1}=G_{3}=...=G_{2N-1}$, $G_{2}=G_{4}=...=G_{2N-2}$, and $G_{1}=G_{3}=...=G_{2N-1}<G_{2}=G_{4}=...=G_{2N-2}$ to map the topologically nontrivial SSH model. Naturally, these conditions need the corresponding parameters to satisfy $|\alpha_{1}|<|\alpha_{2}|<|\alpha_{3}|<...<|\alpha_{N}|$ and $g_{1}>g_{2}>g_{3}>...>g_{N}$, which, further needs the decay of the cavity fields to satisfy $\kappa_{1}>\kappa_{2}>\kappa_{3}>...>\kappa_{N}$. We can always find a set of parameters to satisfy the above conditions. To illustrate it intuitively and simply, we focus on the case of $N=3$, and the corresponding results are shown in Fig.~\ref{fig6}. Obviously, we can easily obtain the condition of $|\alpha_{1}|<|\alpha_{2}|<|\alpha_{3}|$, as shown in Fig~\ref{fig6}(a). And we also have $|G_{1}|=|G_{3}|=|G_{5}|$ and $|G_{2}|=|G_{4}|$ approximately, as shown in Fig.~\ref{fig6}(b). In this way, we can easily expand the number of the unit cell $N$ to a large number, which will directly decrease the effects of the size effect on the topology of the system. Meanwhile, it means that the topologically nontrivial optomechanical SSH model with large scale can be realized to construct the long SSH chain in 1D and SSH network in two dimension (2D).

\subsection{\label{sec.C} Topological phase transition induced via the cavity decay}
\begin{figure}
	\centering
	\subfigure{\includegraphics[width=0.49\linewidth]{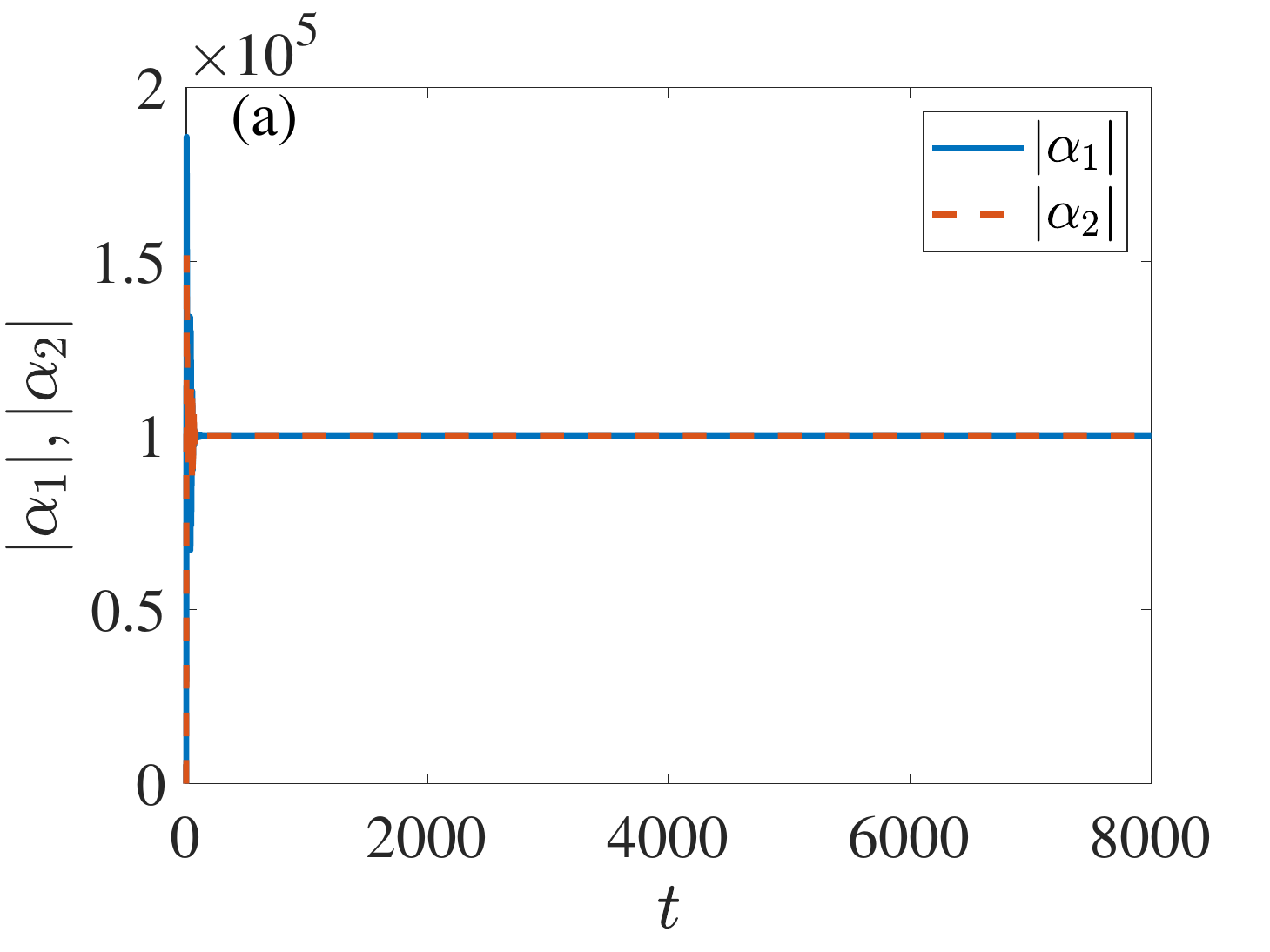}}
	\subfigure{\includegraphics[width=0.49\linewidth]{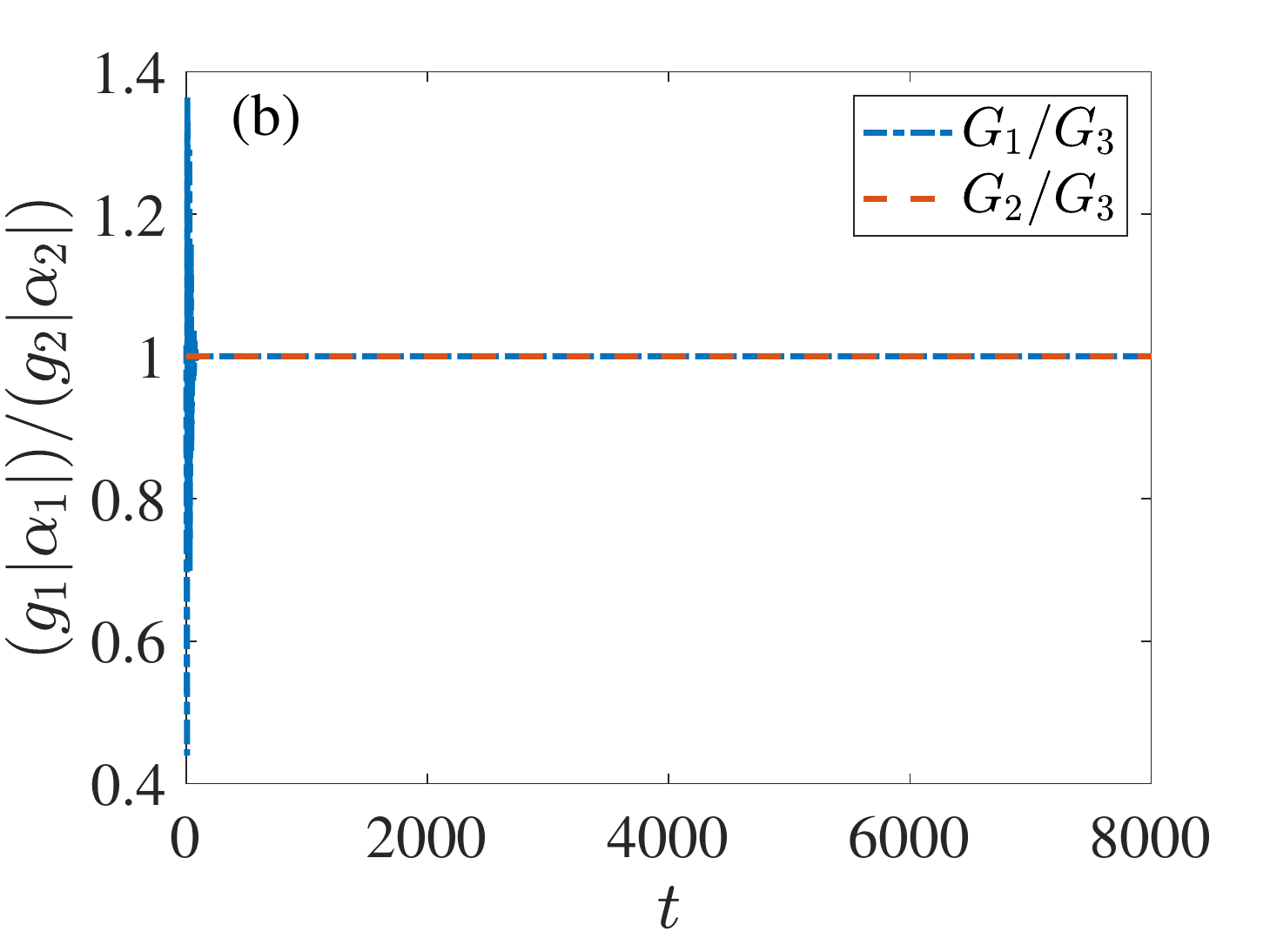}}
	
	\subfigure{\includegraphics[width=0.49\linewidth]{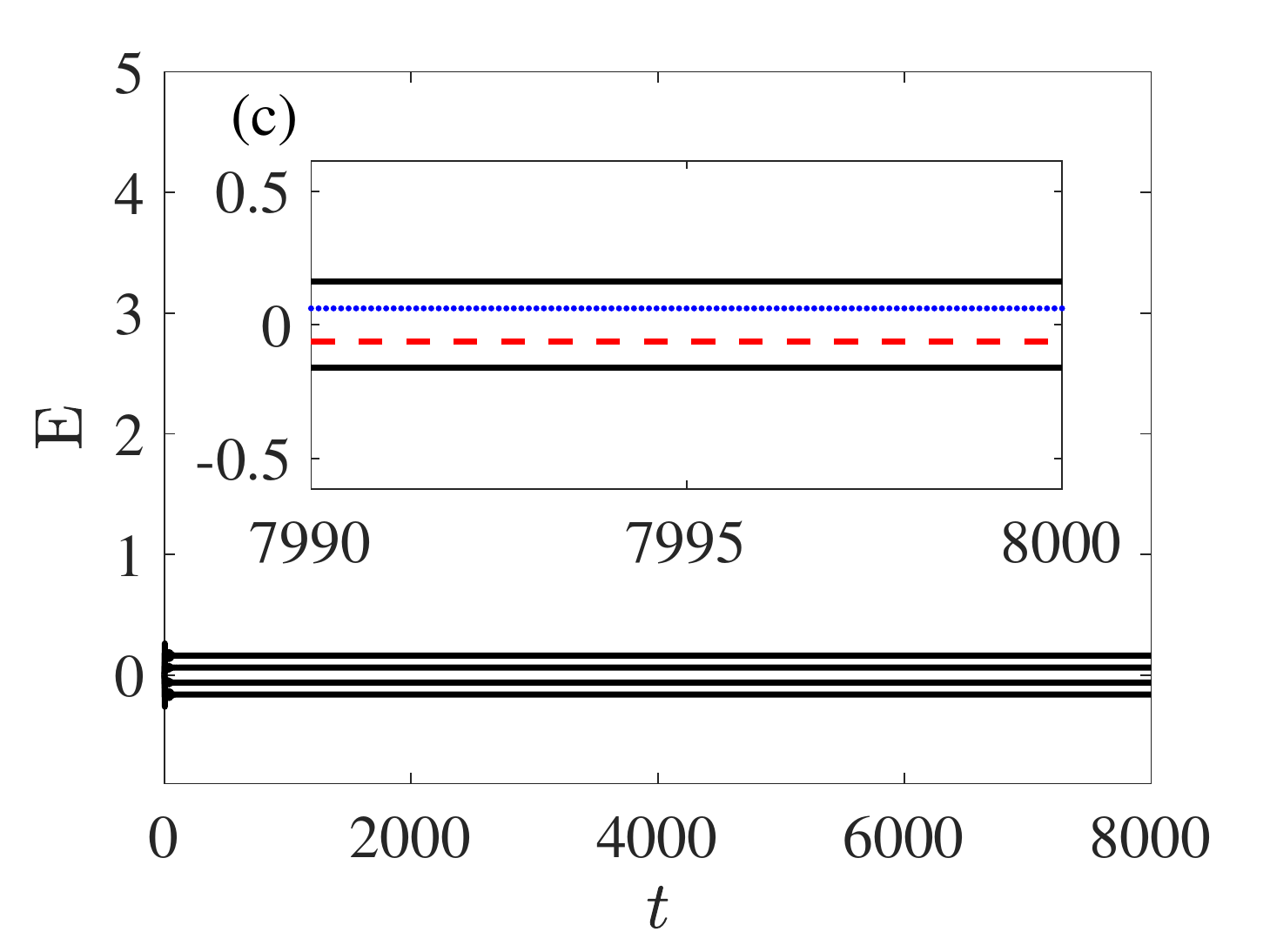}}
	\subfigure{\includegraphics[width=0.49\linewidth]{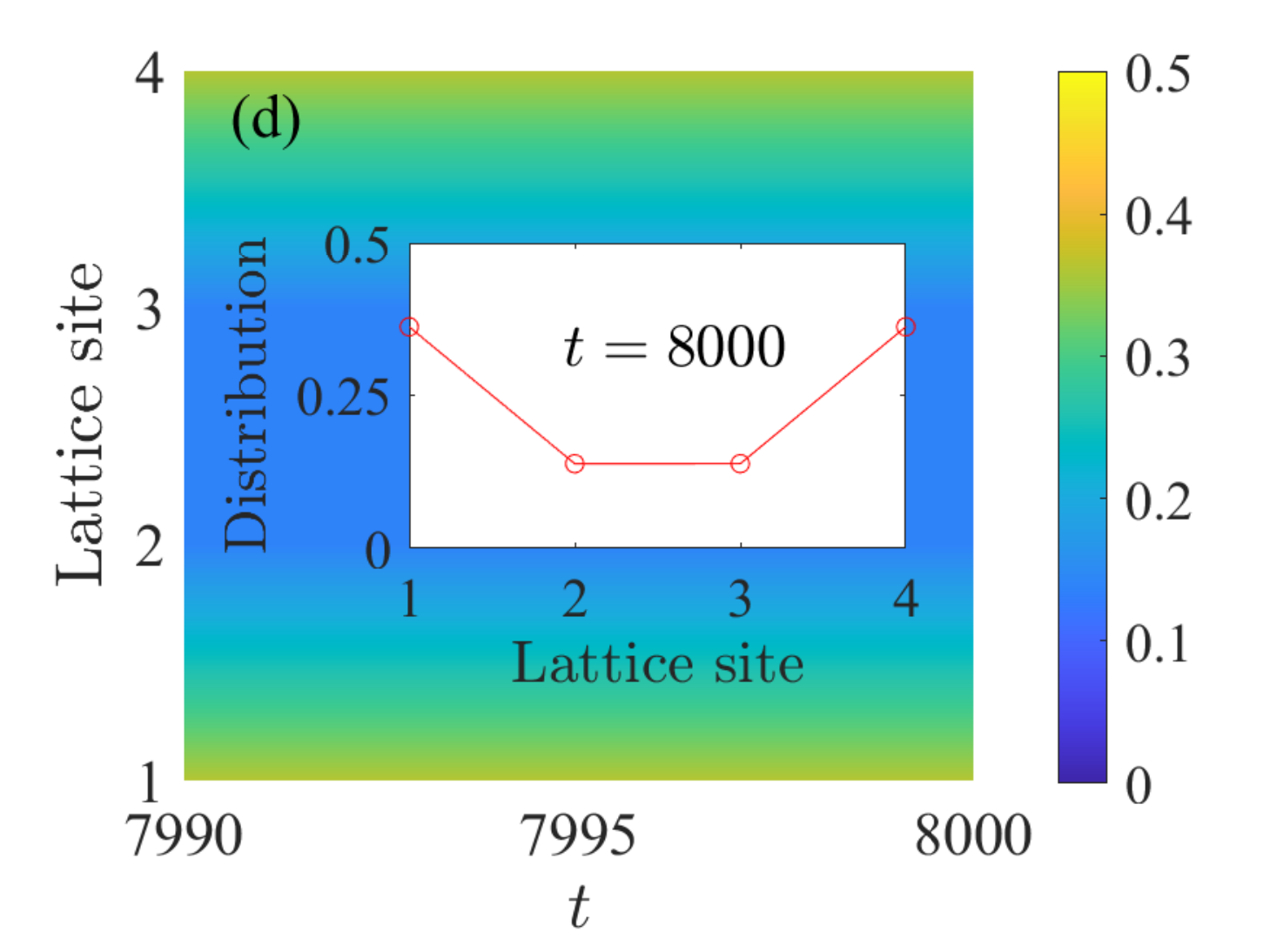}}		
	\caption{The steady state dynamics of the small optomechanical lattice when $|G_{1}|=|G_{2}|=|G_{3}|$. (a) The two final steady cavity fields versus the time $t$. (b) The ratio between the effective optomechanical couplings $G_{1}|$, $|G_{2}|$ and $|G_{3}|$. (c) The energy spectrum of the small optomechanical lattice. (d) The distribution of the blue state in (c). The parameters take $\kappa_{1}=0.1\omega_{b}$ and $\kappa_{2}=0.412\omega_{b}$. Other parameters are the same as the parameters in Figs.~\ref{fig2}(a)-\ref{fig2}(c). We set $\omega_{b}=1$ as the energy unit.}\label{fig7}
\end{figure} 
\begin{figure}
	\centering
	\subfigure{\includegraphics[width=0.49\linewidth]{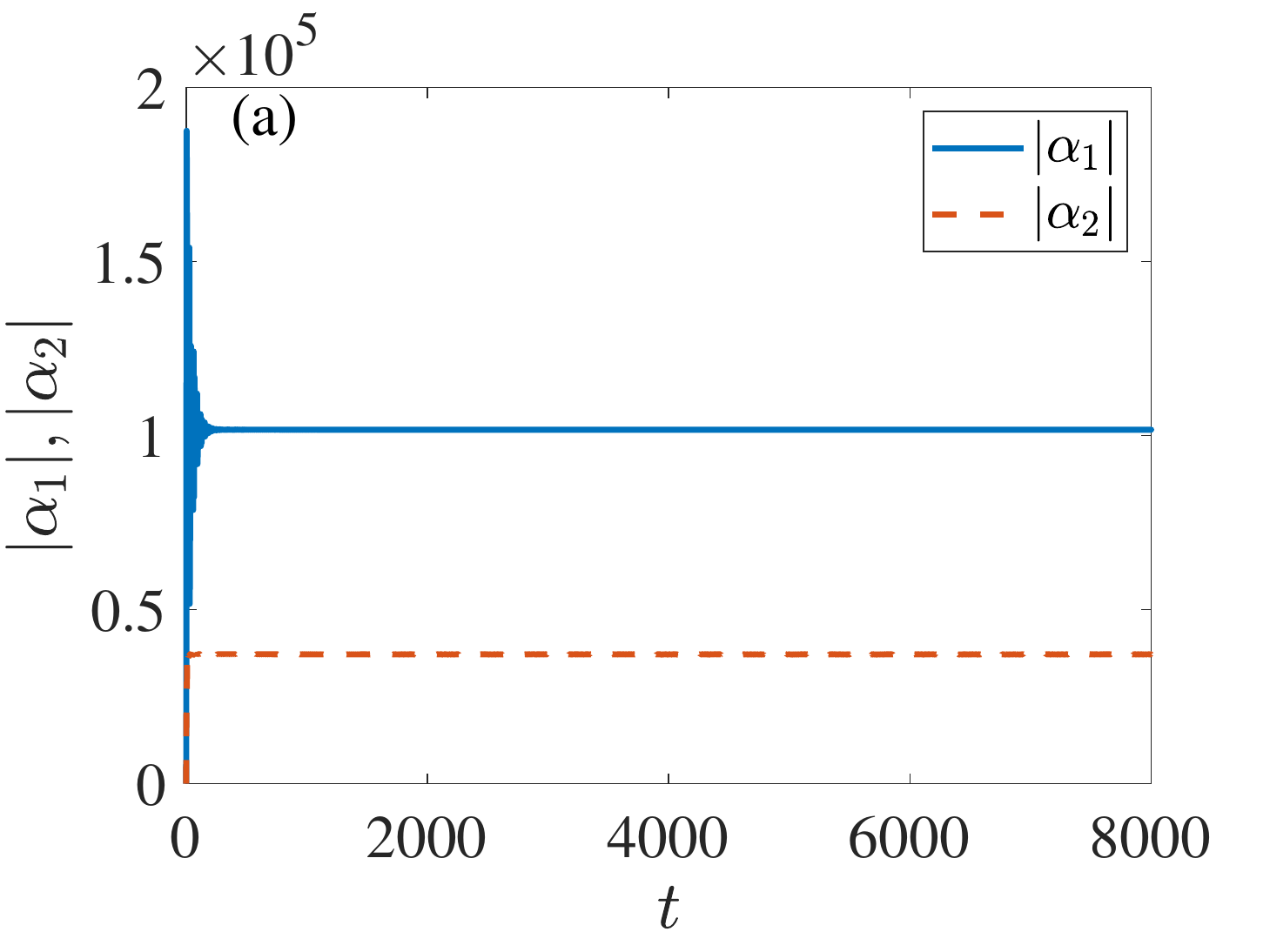}}
	\subfigure{\includegraphics[width=0.49\linewidth]{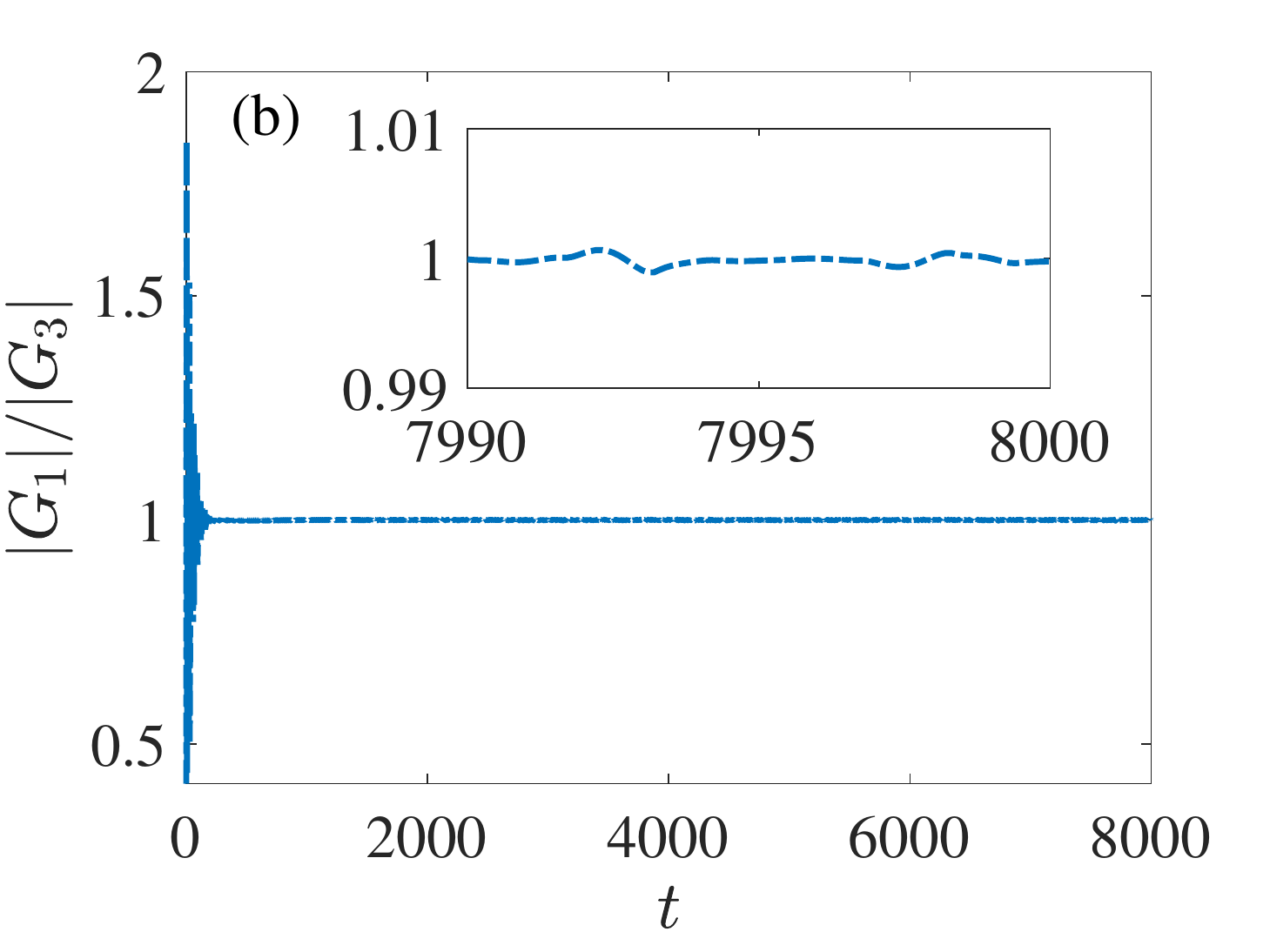}}
	
	\subfigure{\includegraphics[width=0.49\linewidth]{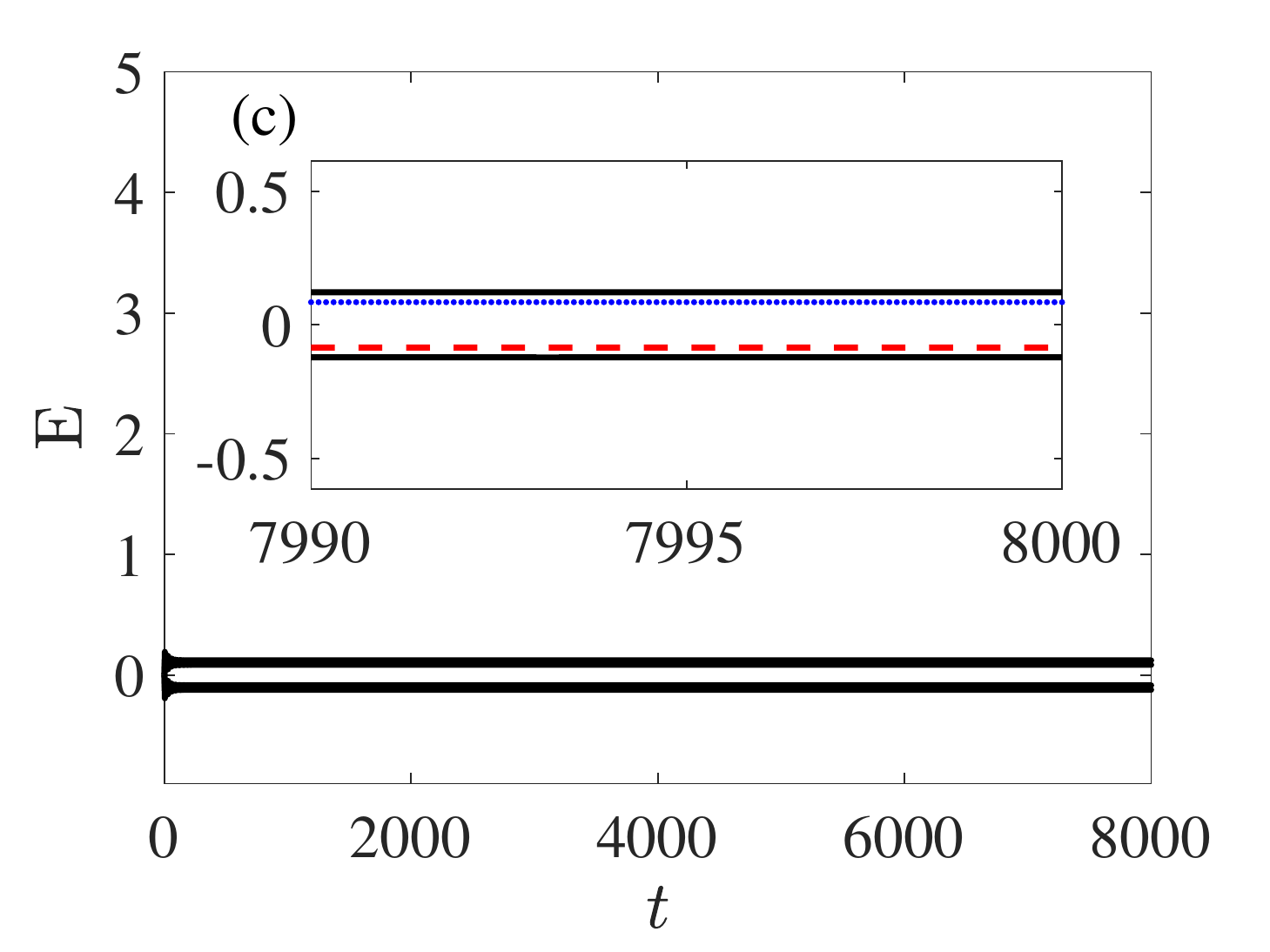}}
	\subfigure{\includegraphics[width=0.49\linewidth]{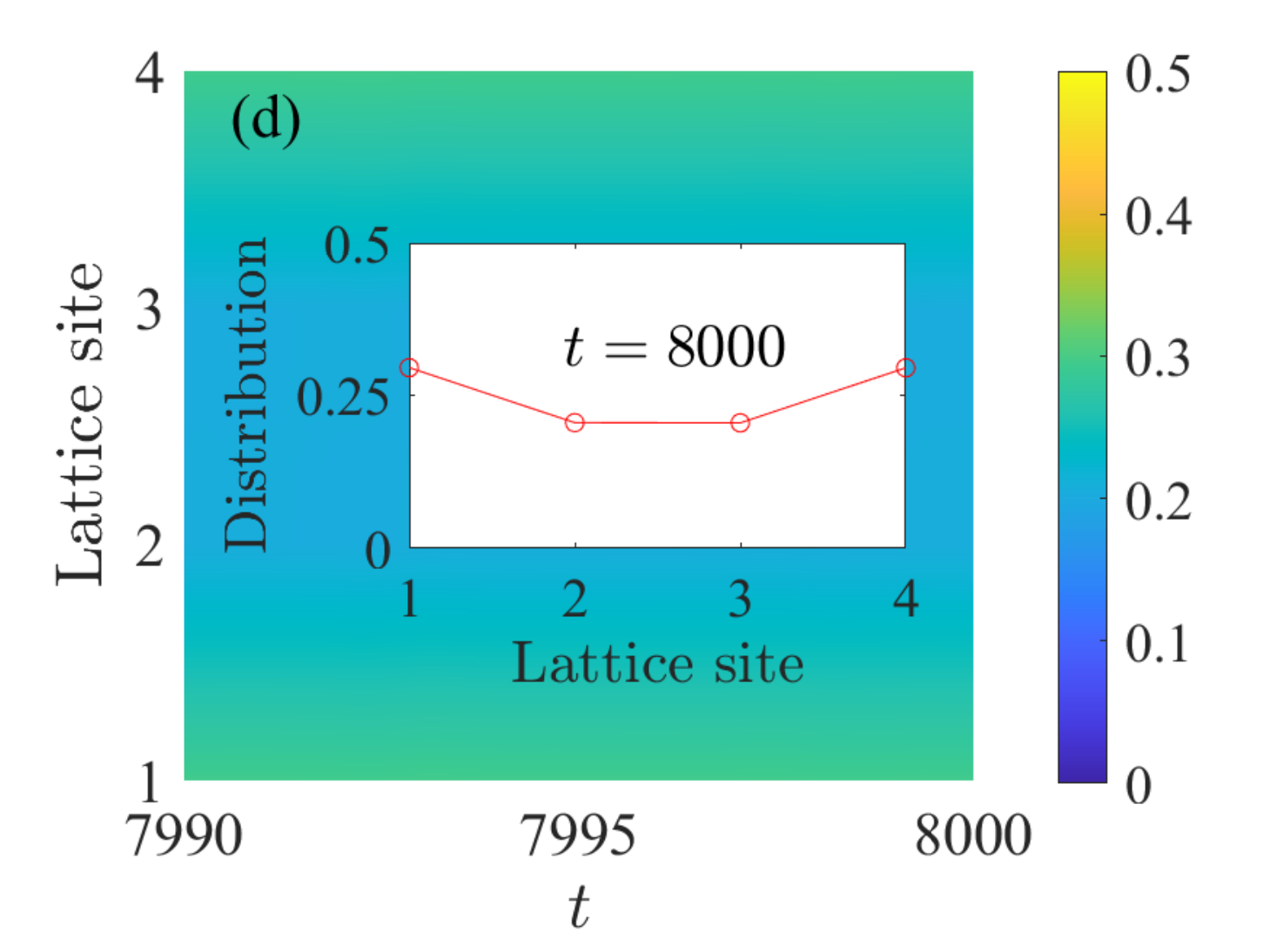}}		
	\caption{The steady state dynamics of the small optomechanical lattice when $|G_{1}|=|G_{3}|\gg|G_{2}|$. (a) The two final steady cavity fields versus the time $t$. (b) The ratio between the effective optomechanical couplings $|G_{1}|$ and $|G_{3}|$. (c) The energy spectrum of the small optomechanical lattice. (d) The distribution of the blue state in (c). The parameters take $g_{1}=1.0\times10^{-6}\omega_{b}$, $g_{2}=2.7375\times10^{-6}\omega_{b}$, $\kappa_{1}=0.1\omega_{b}$ and $\kappa_{2}=5\omega_{b}$. Other parameters are the same as the parameters in Figs.~\ref{fig2}(a)-\ref{fig2}(c). We set $\omega_{b}=1$ as the energy unit.}\label{fig8}
\end{figure}  
All the above discussions are mainly based on the changed decay of the first cavity and the fixed decay of the second cavity, which ensures the nontrivial topology of the small optomechanical lattice. We have revealed that, for a specific cavity field, the larger decay of the cavity field corresponds to a smaller final steady cavity field. Thus, when the decay of the first cavity field is fixed and the decay of the second cavity field is decreased, the final two steady state cavity fields may satisfy $|\alpha_{1}|>|\alpha_{2}|$ ($|G_{1}|>|G_{2}|$), which may further determine the trivial topology. In this way, via designing the decay of the second cavity fields appropriately, we may realize the topological phase transition between the nontrivial SSH phase and the trivial SSH phase.  
             
In Fig.~\ref{fig3}, we have demonstrated that the topological SSH model can be realized when $|\alpha_{1}|=|\alpha_{3}|<|\alpha_{2}|$ (the topological effect is unconspicuous). Based on the conclusions revealed in Fig.~\ref{fig3}, it is easy to find that, with the decay of the second cavity field increasing continually, the final steady-state cavity field $\alpha_{2}$ decreases continually. For example, when $\kappa_{1}=0.1\omega_{b}$ and $\kappa_{2}=0.412\omega_{b}$, we find that the two final steady-state cavity fields now satisfy $|\alpha_{1}|=|\alpha_{2}|$, as shown in Fig.~\ref{fig7}(a). Besides, the optomechanical couplings $g_{1}=g_{2}=1.0\times10^{-6}\omega_{b}$ also ensure that the third effective optomechanical coupling satisfies $\frac{|G_{1}|}{|G_{3}|}=1$, as shown in Fig.~\ref{fig7}(b). Thus, the system always has $|G_{1}|=|G_{2}|=|G_{3}|$, which leads that the system corresponds to a critical point of phase transition since the intra-cell couplings are identical with the inter-cell couplings. To further verify it, we plot the energy spectrum and the distribution of the gap state, as shown in Figs.~\ref{fig7}(c) and \ref{fig7}(d). The numerical results reveal that, compared with the case in Fig.~\ref{fig3}(a), the condition of $|G_{1}|=|G_{2}|=|G_{3}|$ leads that the two originally separated gap states begin to move toward two bulk energy levels, as shown in Fig.~\ref{fig7}(c). At the same time, the distributions of the gap state at two ends sites decrease while the distributions of the gap state at bulk sites increase, as shown in Fig.~\ref{fig7}(d). 

With the decay of the second cavity field further increasing, such as $\kappa_{1}=0.1\omega_{b}$ and $\kappa_{2}=5\omega_{b}$, we find that the second final steady-state cavity field further decreases, leading to $|\alpha_{1}|\gg|\alpha_{2}|$, as shown in Fig.~\ref{fig8}(a). Besides, the optomechanical couplings $g_{1}=1.0\times10^{-6}\omega_{b}$ and $g_{2}=2.7375\times10^{-6}\omega_{b}$ also ensure the third effective optomechanical coupling to satisfy $|G_{1}|=|G_{3}|$, as shown in Fig.~\ref{fig8}(b). In this way, the system always has $|G_{1}|=|G_{3}|\gg|G_{2}|$, which leads the system to be equivalent to a topologically trivial SSH phase since the intra-cell couplings are much larger than the inter-cell couplings. Similarly, we also plot the energy spectrum, as shown in Fig.~\ref{fig8}(c). The numerical results indicate that the two separated gap states further move toward two bulk energy levels and finally integrate into the bulk, leading that the gap states disappear and the system only possesses two bulk energy bands, as shown in Fig.~\ref{fig8}(c). This phenomenon implies that the present small optomechanical lattice is a topological trivial phase. To further clarify it, we plot the distributions of the original gap state, as shown in Fig.~\ref{fig8}(d). The results show that the original gap state now becomes an extended state completely.  

Based on the conclusions mentioned above, it means that, via controlling the decay of the second cavity fields, we can induce the phase transition between the topologically nontrivial SSH phase and the topologically trivial SSH phase through the critical point of phase transition. Our scheme provides a new stage toward the realization of topological phase transition induced by the dispassion based on the small optomechanical lattice.           
\begin{figure}
	\centering
	\includegraphics[width=0.9\linewidth]{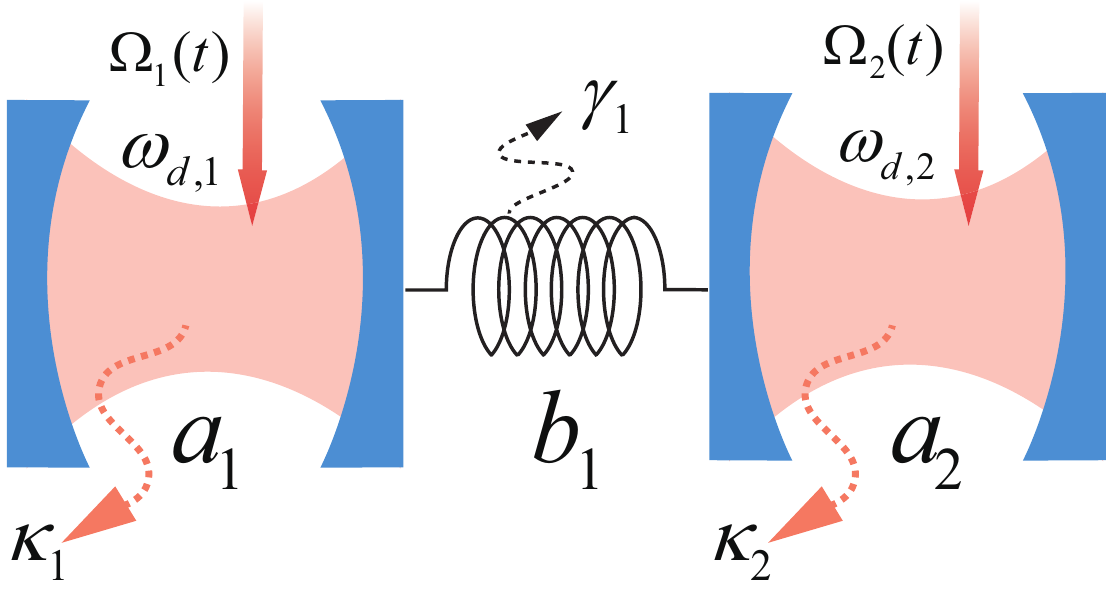}\\
	\caption{Schematic of the 1D small optomechanical lattice with the odd number of lattice sites. The cavity field $a_{1}$ ($a_{2}$) is driven by an external laser with driving amplitude $\Omega_{1}(t)$ ($\Omega_{2}(t)$) and driving frequency $\omega_{d,1}$ ($\omega_{d,2}$). The decay of the cavity field $a_{1}$ ($a_{2}$) and the damping of the resonator $b_{1}$ are $\kappa_{1}$ ($\kappa_{2}$) and $\gamma_{1}$, respectively.}\label{fig9}
\end{figure} 
\begin{figure*}
	\centering
	\subfigure{\includegraphics[width=0.32\linewidth]{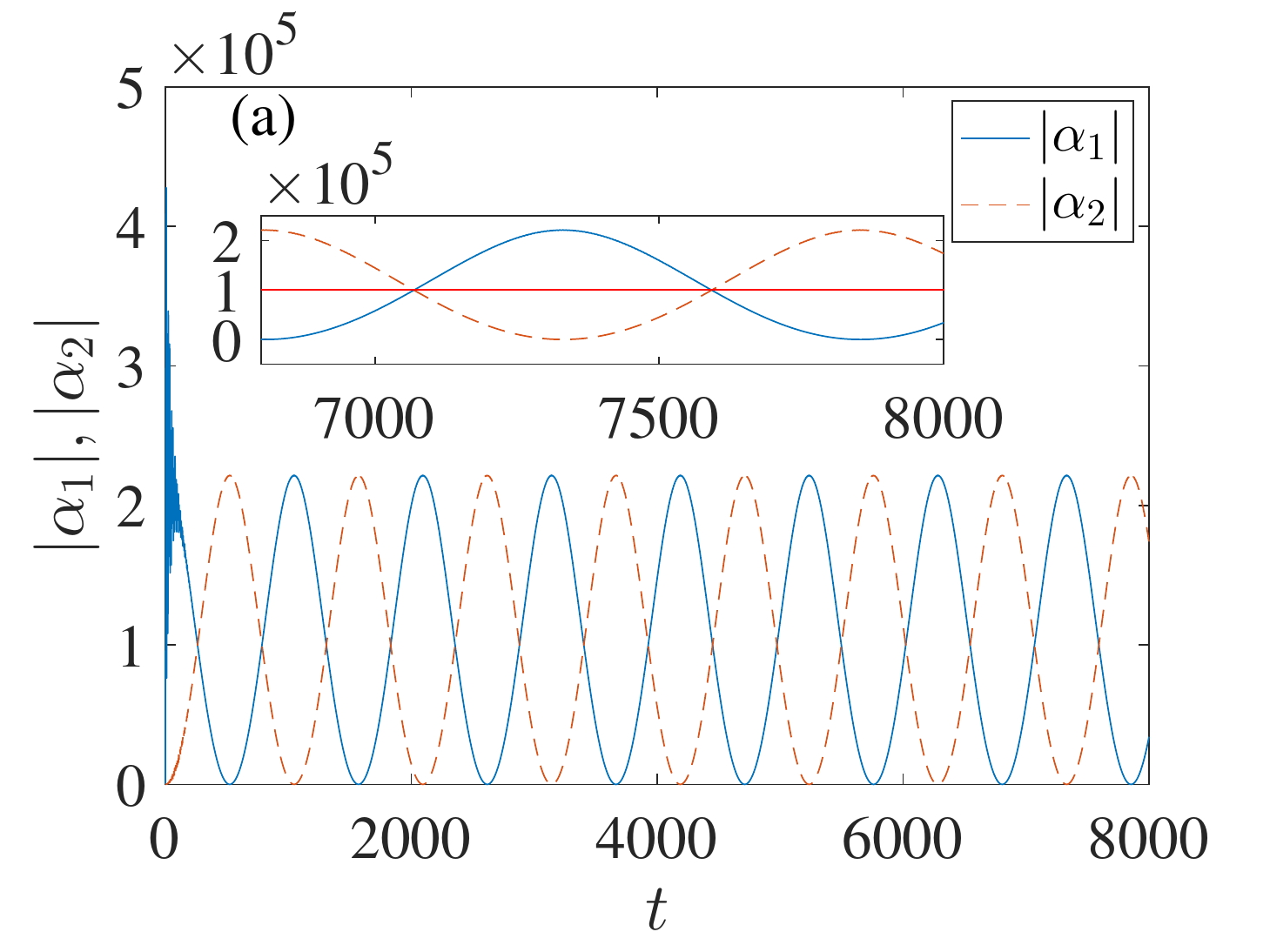}}
	\subfigure{\includegraphics[width=0.32\linewidth]{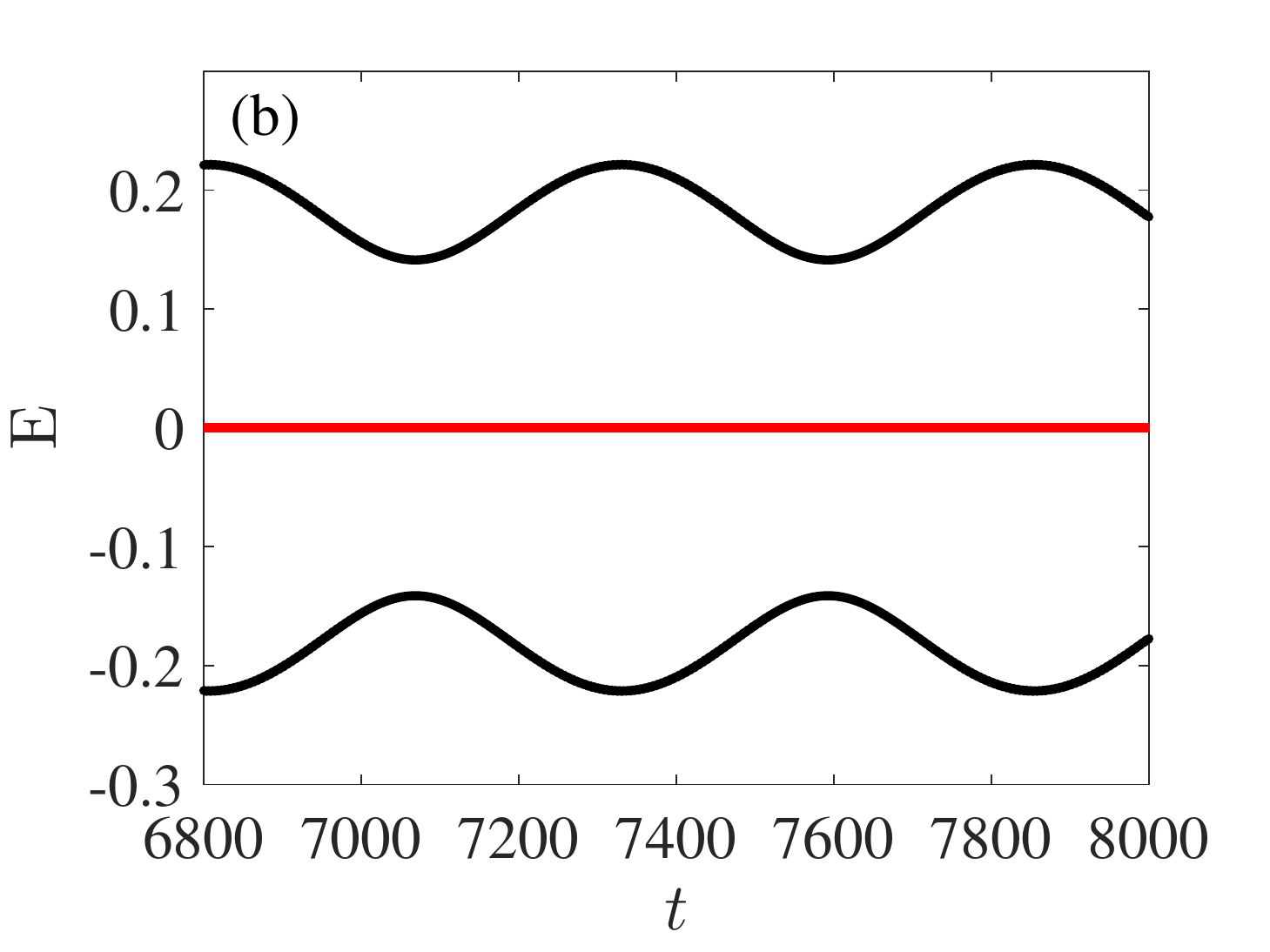}}	
	\subfigure{\includegraphics[width=0.32\linewidth]{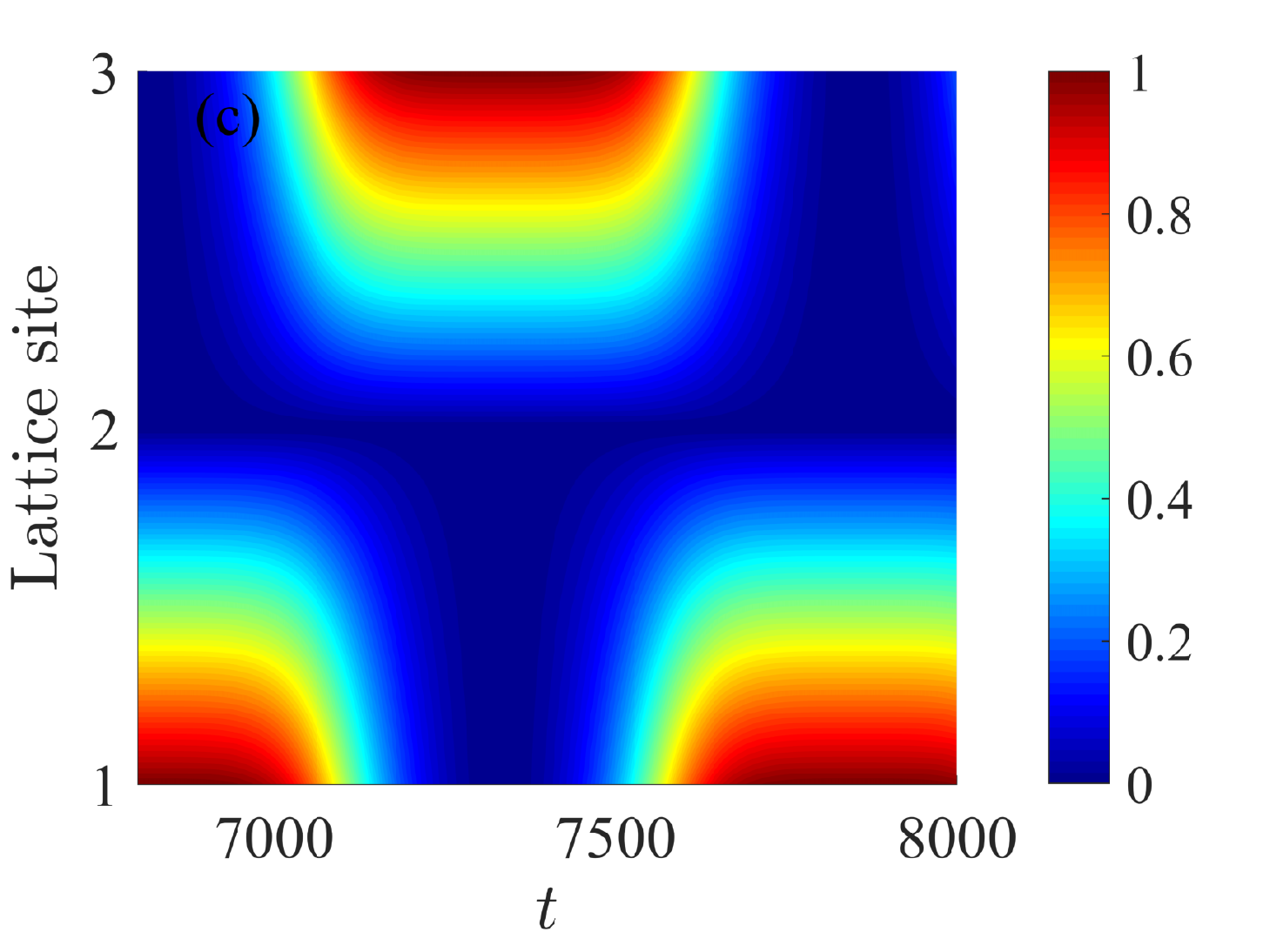}}	
	\caption{The steady-state cavity fields, energy spectrum, and the distribution of the gap state in the small optomechanical lattice. (a) The distributions of the steady-state cavity fields $\alpha_{1}$ and $\alpha_{2}$. (b) The energy spectrum of the the small optomechanical lattice with three lattice sites. (c) The distribution of the zero energy mode with the time $t$. The parameters take $\omega_{b,1}=\omega_{b}$, $\Delta_{a,1}=\Delta_{a,2}=\omega_{b}$, $g_{1}=1\times10^{-6}\omega_{b}$, $\Omega=1\times10^{5}\omega_{b}$, $\nu_{1}=\nu_{2}=0.006\omega_{b}$, $\kappa_{1}=\kappa_{2}=0.1\omega_{b}$, and $\gamma_{1}=\gamma_{2}=1\times10^{-5}\omega_{b}$. We set $\omega_{b}=1$ as the energy unit.}\label{fig10}
\end{figure*}

\section{\label{sec.4} Photonic topological state transfer based on a small optomechanical lattice with periodical driving}
The optomechanical lattice provides a new kind of platform to investigate the SSH model with nontrivial topological edge states. The nontrivial edge states have many potential applications in quantum information processing and quantum computation, such as, the robust quantum state transfer assisted by the edge channel~\cite{dlaska2017robust,mei2018robust,qi2020controllable}. Note that, in Ref.~\cite{qi2020controllable}, we have proposed a controllable photonic and phononic topological state transfer scheme, in which the realization of the state transfer depends on the assumption of the periodically modulated effective optomechanical coupling. However, the insightful mechanism of the modulated terms, under the realistic parameters regime, is not fully revealed. Here, to reveal the insightful mechanism, we propose a small optomechanical lattice composed by two cavity fields and one resonator, in which the two cavity fields are driven via two periodically time-dependent external laser, as shown in Fig.~\ref{fig9}. 

We stress that we mainly aim to reveal the physical feasibility of the modulated effective optomechanical coupling $G_{n}$ under the realistic parameters regime, rather than the discussions of large-scale scalability. Thus, once the physical feasibility of the modulated effective optomechanical coupling $G_{n}$ based on the small lattice, under the realistic parameters regime, is fully demonstrated, the large-scale optomechanical lattice can be naturally extended. 

The present small optomechanical lattice, besides can keep its the simplest form in space, also has the following significant advantages. As revealed in Ref.~\cite{mei2018robust}, the topological state transfer is protected by the energy gap, meaning that the larger energy gap ensures the system to be more robust to the disorder and perturbation. Note that the SSH model with small size exhibits a larger energy gap compared with the SSH model with the large size, implying the strengthened topological protection, namely, the state transfer in the small optomechanical lattice is more immune to the disorder and perturbation in experiment. Moreover, the topological state transfer needs that the varying of the parameter is much smaller than the energy gap for the satisfaction of the adiabatic evolution~\cite{mei2018robust}. Naturally, for the smaller optomechanical lattice with the larger energy gap, the varying of the parameter can be relatively faster compared with the optomechanical lattice with the larger size. In this way, the adiabatic condition can be realized much easier in experiment. 

Thus, in the following, we focus on the optomechanical lattice with three sites to clarify the feasibility of the state transfer. For example, we take the two cavity fields to be driven by two time-dependent lasers with amplitudes $\Omega_{1}(t)=\Omega(1-\cos\nu_{1}t)$ and $\Omega_{2}(t)=\Omega(1+\cos\nu_{2}t)$. Then, the effective steady-state dynamic can be dominated by the following Hamiltonian, with
\begin{eqnarray}\label{e07}
H&=&-g_{1}\alpha_{1}(a_{1}^{\dag}b_{1}+b_{1}^{\dag}a_{1})+g_{1}\alpha_{2}(a_{2}^{\dag}b_{1}+b_{1}^{\dag}a_{2}),
\end{eqnarray}  
where $g_{1}$ is the optomechanical coupling between the two cavity fields and the resonator, $\alpha_{1}$ and $\alpha_{2}$ are the final steady-state cavity fields. The final steady-state cavity fields $\alpha_{1}$ and $\alpha_{2}$ can be described by the following differential equations, with
\begin{eqnarray}\label{e08}
\dot{\alpha_{1}}&=&-i[\Delta_{a,1}-g_{1}(\beta_{1}+\beta_{1}^{\ast})]\alpha_{1}-i\Omega_{1}(t)-\frac{\kappa_{1}}{2}\alpha_{1},\cr\cr
\dot{\beta_{1}}&=&-i(\omega_{b,1}\beta_{1}-g_{1}|\alpha_{1}|^{2})-\frac{\gamma_{1}}{2}\beta_{1},\cr\cr
\dot{\alpha_{2}}&=&-i[\Delta_{a,2}+g_{1}(\beta_{1}+\beta_{1}^{\ast})]\alpha_{2}-i\Omega_{2}(t)-\frac{\kappa_{2}}{2}\alpha_{2},
\end{eqnarray} 
where $\Delta_{a,1}$ ($\Delta_{a,2}$) is the cavity field detuning, $\kappa_{1}$ ($\kappa_{2}$) is the decay of the first (second) cavity field, and $\gamma_{1}$ is the damping of the resonator. According to Floquet’s theory~\cite{barone1977floquet,Berdanier2017floquet}, the final steady-state cavity fields will have the same period of
the driving amplitudes in the long-time limit, which means that the effective optomechanical couplings $-g_{1}\alpha_{1}$ and $g_{1}\alpha_{2}$ should also be periodically time-dependent. It indicates that we can realize the mapping of modulated topological SSH model with odd number of lattice sites if we regard the cavity field $a_{1}$ ($a_{2}$) and the resonator $b_{1}$ as the sites of the optomechanical lattice. To further clarify it, we plot the final steady-state cavity fields $\alpha_{1}$ and $\alpha_{2}$ versus the time $t$ when $\nu_{1}=\nu_{2}=\nu=0.006\omega_{b}$, as shown in Fig.~\ref{fig10}(a). The numerical results reveal that the final steady-state cavity fields $\alpha_{1}$ and $\alpha_{2}$ are indeed periodically time-dependent and have the same period as the time-dependent driving amplitudes $\Omega_{1}(t)$ and $\Omega_{2}(t)$. Thus, we have steady-state cavity fields $\alpha_{1}\approx(1-\cos\nu t)\times10^{5}\omega_{b}$ and $\alpha_{2}\approx(1+\cos\nu t)\times10^{5}\omega_{b}$, which leads that the effective optomechanical couplings satisfy $-g_{1}\alpha_{1}\approx-0.1(1-\cos\nu t)\omega_{b}$ and $g_{1}\alpha_{2}\approx0.1(1+\cos\nu t)\omega_{b}$. Thus, the effective tight-binding Hamiltonian can be written as (in the unit of $\omega_{b}$)
\begin{eqnarray}\label{e09}
H&=&-0.1(1-\cos\nu t)(a_{1}^{\dag}b_{1}+b_{1}^{\dag}a_{1})\cr\cr
&&+0.1(1+\cos\nu t)(a_{2}^{\dag}b_{1}+b_{1}^{\dag}a_{2}).
\end{eqnarray} 
   
Obviously, the above effective Hamiltonian is equivalent to a periodical-modulated SSH model, in which the parameter $\nu t$ is equivalent to the periodic parameter $\theta$ in Ref.~\cite{qi2020controllable}. To further illuminate it, we simulate the energy spectrum numerically via using the effective optomechanical couplings $-g_{1}\alpha_{1}$ and $g_{1}\alpha_{2}$, as shown in Fig.~\ref{fig10}(b). The numerical results reveal that the present small optomechanical lattice always has a zero energy mode locating in the gap, which means that the present small optomechanical lattice with periodical driving can indeed be mapped into a periodically modulated SSH model with the odd number of lattice sites. The periodically modulated SSH model  with the odd number of lattice sites is widely used to implement the quantum state transfer~\cite{dlaska2017robust,mei2018robust,qi2020controllable} via the topological zero energy mode. We simulate the distribution of the zero energy state with the developing of the time $t$, as shown in Fig.~\ref{fig10}(c). Obviously, the zero energy mode is localized at the first cavity field and the last cavity field alternately with the developing of the time $t$, which implies that we can realize the state transfer between the two cavity fields via the zero energy mode. As mentioned above, the system needs to satisfy the adiabatic evolution condition, which means that the parameter $\nu$ needs to be less than the energy gap to satisfy the adiabatic evolution condition~\cite{mei2018robust} in the state transfer scheme. As shown in Fig.~\ref{fig10} (b), the minimal energy gap of the system is about $0.2\omega_{b}$, which is much larger than parameter $\nu$ ($\nu=0.006\omega_{b}$). It means that the state transfer scheme assisted by topological zero energy mode can be achieved under the present parameters regime.     

We stress that, all the above discussions are based on a small optomechanical lattice with three lattice sites. Meanwhile, our scheme of controllable photonic topological state transfers can also be extended to a large scale of the optomechanical lattice relying on the feasibility of the manipulation at single site level. In this way, the large scale of quantum information processing can be constructed based on the optomechanical lattice, which greatly extends the applications of the topological optomechanical system in quantum information processing.

\section{\label{sec.5}Conclusions}
In conclusion, we have proposed a scheme to investigate the topological SSH phase transition and photonic topological state transfer based on the small optomechanical lattice. After deriving the effective steady-state Hamiltonian, we find that the small optomechanical lattice can be mapped into a tight-binding model. We reveal that, the system can be equivalent to a topologically nontrivial SSH phase by choosing the appropriate effective optomechanical coupling. Note that, the excepted topological phase transition can be obtained via controlling the decay of the cavity field and the optomechanical coupling, in which the system experiences the phase transition between the topologically nontrivial SSH phase and topologically trivial SSH phase through the critical point of phase transition. Besides, we also investigate the photonic topological state transfer based on the small optomechanical lattice when the cavity fields are driven by the external lasers with different periodically time-dependent amplitudes. Our scheme provides the new and the possible optical platform to induce the topological phase transition and realize the topologically protected state transfer.

\begin{acknowledgments}
This work was supported by the National Natural Science Foundation of China under Grant Nos.
61822114, 11874132, 61575055, and 11575048.
\end{acknowledgments}




\begin{thebibliography}{10}
	\newcommand{\enquote}[1]{``#1''}
	
	\bibitem{hasan2010colloquium}
	M.~Z. Hasan and C
	.~L. Kane, Colloquium: topological insulators, Rev. Mod. Phys.
	\textbf{82}, 3045 (2010).
	
	\bibitem{qi2011topological}
	X.~L. Qi and S.~C. Zhang, Topological insulators and superconductors, Rev. Mod.
	Phys. \textbf{83}, 1057 (2011).
	
	\bibitem{chiu2016classification}
	C.~K. Chiu, J.~C. Teo, A.~P. Schnyder, and S.~Ryu, Classification of
	topological quantum matter with symmetries, Rev. Mod. Phys. \textbf{88},
	035005 (2016).
	
	\bibitem{bansil2016colloquium}
	A.~Bansil, H.~Lin, and T.~Das, Colloquium: Topological band theory, Rev. Mod.
	Phys. \textbf{88}, 021004 (2016).
	
	\bibitem{dlaska2017robust}
	C.~Dlaska, B.~Vermersch, and P.~Zoller, Robust quantum state transfer via
	topologically protected edge channels in dipolar arrays, Quantum Sci. Technol. \textbf{2}, 015001 (2017).
	
	\bibitem{mei2018robust}
	F.~Mei, G.~Chen, L.~Tian, S.~L.~Zhu, and S.~Jia, Robust quantum state transfer via topological edge states in superconducting qubit chains, Phys. Rev. A \textbf{98}, 012331 (2018)
	
	\bibitem{qi2020controllable}
	L.~Qi, G.~L. Wang, S. Liu, S. Zhang, and H.~F. Wang, Controllable double quantum state transfers by one topological channel in a frequency-modulated optomechanical array, Opt. Lett. \textbf{45}, 2018-2021 (2020)
	
	\bibitem{aasen2016milestones}
	D.~Aasen, M.~Hell, R.~V. Mishmash, A.~Higginbotham, J.~Danon, M.~Leijnse, T.~S.
	Jespersen, J.~A. Folk, C.~M. Marcus, K.~Flensberg \emph{et~al.}, Milestones
	toward majorana-based quantum computing, Phys. Rev. X \textbf{6}, 031016
	(2016).
	
	\bibitem{Sarma2015}
	S.~D. Sarma, M.~Freedman, and C.~Nayak, Majorana zero modes and topological
	quantum computation, npj Quantum Inf. \textbf{1}, 15001 (2015).
	
	\bibitem{harari2018topological}
	G.~Harari, M.~A. Bandres, Y.~Lumer, M.~C. Rechtsman, Y.~D. Chong,
	M.~Khajavikhan, D.~N. Christodoulides, and M.~Segev, Topological insulator
	laser: theory, Science \textbf{359}, eaar4003 (2018).
	
	\bibitem{bandres2018topological}
	M.~A. Bandres, S.~Wittek, G.~Harari, M.~Parto, J.~Ren, M.~Segev, D.~N.
	Christodoulides, and M.~Khajavikhan, Topological insulator laser:
	Experiments, Science \textbf{359}, eaar4005 (2018).
	
	\bibitem{khanikaev2013photonic}
	A.~B. Khanikaev, S.~H. Mousavi, W.~K. Tse, M.~Kargarian, A.~H. MacDonald, and
	G.~Shvets, Photonic topological insulators, Nat. Mater. \textbf{12}, 233--239
	(2013).
	
	\bibitem{rechtsman2013photonic}
	M.~C. Rechtsman, J.~M. Zeuner, Y.~Plotnik, Y.~Lumer, D.~Podolsky, F.~Dreisow,
	S.~Nolte, M.~Segev, and A.~Szameit, Photonic floquet topological insulators,
	Nature \textbf{496}, 196--200 (2013).
	
	\bibitem{yan2014topological}
	P.~Yan, R.~Lin, H.~Chen, H.~Zhang, A.~Liu, H.~Yang, and S.~Ruan, Topological
	insulator solution filled in photonic crystal fiber for passive mode-locked
	fiber laser, IEEE Photon. Technol. Lett. \textbf{27}, 264--267 (2014).
	
	\bibitem{lu2016symmetry}
	L.~Lu, C.~Fang, L.~Fu, S.~G. Johnson, J.~D. Joannopoulos, and
	M.~Solja{\v{c}}i{\'c}, Symmetry-protected topological photonic crystal in
	three dimensions, Nat. Phys. \textbf{12}, 337--340 (2016).
	
	\bibitem{gao2015stable}
	L.~Gao, T.~Zhu, W.~Huang, and Z.~Luo, Stable, ultrafast pulse mode-locked by
	topological insulator bi2se3 nanosheets interacting with photonic crystal
	fiber: From anomalous dispersion to normal dispersion, IEEE Photon. J.
	\textbf{7}, 1--8 (2015).
	
	\bibitem{shalaev2019robust}
	M.~I. Shalaev, W.~Walasik, A.~Tsukernik, Y.~Xu, and N.~M. Litchinitser, Robust
	topologically protected transport in photonic crystals at telecommunication
	wavelengths, Nat. Nanotechnol. \textbf{14}, 31--34 (2019).
	
	\bibitem{skirlo2015experimental}
	S.~A. Skirlo, L.~Lu, Y.~Igarashi, Q.~Yan, J.~Joannopoulos, and
	M.~Solja{\v{c}}i{\'c}, Experimental observation of large chern numbers in
	photonic crystals, Phys. Rev. Lett. \textbf{115}, 253901 (2015).
	
	\bibitem{xu2016accidental}
	L.~Xu, H.~X. Wang, Y.~D. Xu, H.~Y. Chen, and J.~H. Jiang, Accidental degeneracy
	in photonic bands and topological phase transitions in two-dimensional
	core-shell dielectric photonic crystals, Opt. Express \textbf{24},
	18059--18071 (2016).
	
	\bibitem{wu2015scheme}
	L.~H. Wu and X.~Hu, Scheme for achieving a topological photonic crystal by
	using dielectric material, Phys. Rev. Lett. \textbf{114}, 223901 (2015).
	
	\bibitem{tomita1986observation}
	A.~Tomita and R.~Y. Chiao, Observation of berry's topological phase by use of
	an optical fiber, Phys. Rev. Lett. \textbf{57}, 937 (1986).
	
	\bibitem{chen2014experimental}
	W.~J. Chen, S.~J. Jiang, X.~D. Chen, B.~Zhu, L.~Zhou, J.~W. Dong, and C.~T.
	Chan, Experimental realization of photonic topological insulator in a
	uniaxial metacrystal waveguide, Nat. Commun. \textbf{5}, 1--7 (2014).
	
	\bibitem{ke2019topological}
	S.~Ke, D.~Zhao, J.~Liu, Q.~Liu, Q.~Liao, B.~Wang, and P.~Lu, Topological bound
	modes in anti-pt-symmetric optical waveguide arrays, Opt. Express
	\textbf{27}, 13858--13870 (2019).
	
	\bibitem{longhi2013zak}
	S.~Longhi, Zak phase of photons in optical waveguide lattices, Opt. Lett.
	\textbf{38}, 3716--3719 (2013).
	
	\bibitem{blanco2016topological}
	A.~Blanco-Redondo, I.~Andonegui, M.~J. Collins, G.~Harari, Y.~Lumer, M.~C.
	Rechtsman, B.~J. Eggleton, and M.~Segev, Topological optical waveguiding in
	silicon and the transition between topological and trivial defect states,
	Phys. Rev. Lett. \textbf{116}, 163901 (2016).
	
	\bibitem{liang2013optical}
	G.~Liang and Y.~Chong, Optical resonator analog of a two-dimensional
	topological insulator, Phys. Rev. Lett. \textbf{110}, 203904 (2013).
	
	\bibitem{mivehvar2017superradiant}
	F.~Mivehvar, H.~Ritsch, and F.~Piazza, Superradiant topological peierls
	insulator inside an optical cavity, Phys. Rev. Lett. \textbf{118}, 073602
	(2017).
	
	\bibitem{liang2014optical}
	G.~Liang and Y.~Chong, Optical resonator analog of a photonic topological
	insulator: a finite-difference time-domain study, International Journal of
	Modern Physics B \textbf{28}, 1441007 (2014).
	
	\bibitem{poli2015selective}
	C.~Poli, M.~Bellec, U.~Kuhl, F.~Mortessagne, and H.~Schomerus, Selective
	enhancement of topologically induced interface states in a dielectric
	resonator chain, Nat. Commun. \textbf{6}, 1--5 (2015).
	
	\bibitem{he2016topological}
	C.~He, Z.~Li, X.~Ni, X.~C. Sun, S.~Y. Yu, M.~H. Lu, X.~P. Liu, and Y.~F. Chen,
	Topological phononic states of underwater sound based on coupled ring
	resonators, Appl. Phys. Lett. \textbf{108}, 031904 (2016).
	
	\bibitem{leykam2018reconfigurable}
	D.~Leykam, S.~Mittal, M.~Hafezi, and Y.~D. Chong, Reconfigurable topological
	phases in next-nearest-neighbor coupled resonator lattices, Phys. Rev. Lett.
	\textbf{121}, 023901 (2018).
	
	\bibitem{tangpanitanon2016topological}
	J.~Tangpanitanon, V.~M. Bastidas, S.~Al-Assam, P.~Roushan, D.~Jaksch, and D.~G.
	Angelakis, Topological pumping of photons in nonlinear resonator arrays,
	Phys. Rev. Lett. \textbf{117}, 213603 (2016).
	
	\bibitem{hafezi2013imaging}
	M.~Hafezi, S.~Mittal, J.~Fan, A.~Migdall, and J.~Taylor, Imaging topological
	edge states in silicon photonics, Nat. Photonics \textbf{7}, 1001--1005
	(2013).
	
	\bibitem{ma2016all}
	T.~Ma and G.~Shvets, All-si valley-hall photonic topological insulator, New J.
	Phys. \textbf{18}, 025012 (2016).
	
	\bibitem{he2019silicon}
	X.~T. He, E.~T. Liang, J.~J. Yuan, H.~Y. Qiu, X.~D. Chen, F.~L. Zhao, and J.~W.
	Dong, A silicon-on-insulator slab for topological valley transport, Nat.
	Commun. \textbf{10}, 1--9 (2019).
	
	\bibitem{mei2016witnessing}
	F.~Mei, Z.~Y. Xue, D.~W. Zhang, L.~Tian, C.~Lee, and S.~L. Zhu, Witnessing
	topological weyl semimetal phase in a minimal circuit-QED lattice, Quantum Sci. Technol. \textbf{1}, 015006 (2016).
	
	\bibitem{huang2016realizing}
	Y.~Huang, Z.~Q. Yin, and W.~Yang, Realizing a topological transition in a
	non-hermitian quantum walk with circuit QED, Phys. Rev. A \textbf{94}, 022302
	(2016).
	
	\bibitem{mei2015simulation}
	F.~Mei, J.~B. You, W.~Nie, R.~Fazio, S.~L. Zhu, and L.~C. Kwek, Simulation and
	detection of photonic chern insulators in a one-dimensional circuit-QED
	lattice, Phys. Rev. A \textbf{92}, 041805 (2015).
	
	\bibitem{qi2018simulation}
	L.~Qi, Y.~Xing, J.~Cao, X.-X. Jiang, C.~S. An, A.~D. Zhu, S.~Zhang, and H.~F.
	Wang, Simulation and detection of the topological properties of a modulated
	rice-mele model in a one-dimensional circuit-QED lattice, Sci. China Phys. Mech. Astron. \textbf{61}, 080313 (2018).
	
	\bibitem{tan2019simulation}
	X.~Tan, Y.~Zhao, Q.~Liu, G.~Xue, H.~F. Yu, Z.~Wang, and Y.~Yu, Simulation and
	manipulation of tunable weyl-semimetal bands using superconducting quantum
	circuits, Phys. Rev. Lett. \textbf{122}, 010501 (2019).
	
	\bibitem{aspelmeyer2014cavity}
	M.~Aspelmeyer, T.~J. Kippenberg, and F.~Marquardt, Cavity optomechanics, Rev.
	Mod. Phys. \textbf{86}, 1391 (2014).
	
	\bibitem{kippenberg2008cavity}
	T.~J. Kippenberg and K.~J. Vahala, Cavity optomechanics: back-action at the
	mesoscale, Science \textbf{321}, 1172--1176 (2008).
	
	\bibitem{eichenfield2009optomechanical}
	M.~Eichenfield, J.~Chan, R.~M. Camacho, K.~J. Vahala, and O.~Painter,
	Optomechanical crystals, Nature \textbf{462}, 78--82 (2009).
	
	\bibitem{kippenberg2007cavity}
	T.~J. Kippenberg and K.~J. Vahala, Cavity opto-mechanics, Opt. Express
	\textbf{15}, 17172--17205 (2007).
	
	\bibitem{wang2018normal}
    T.~Wang, M.~H. Zheng, C.~H. Bai, D.~Y. Wang, A.~D. Zhu, H.~F. Wang, and
    S.~Zhang, Normal-mode splitting and optomechanically induced absorption,
    amplification, and transparency in a hybrid optomechanical system, Ann. Phys. (Berlin) \textbf{530}, 1800228 (2018).	
	
	\bibitem{dobrindt2008parametric}
	J.~M. Dobrindt, I.~Wilson-Rae, and T.~J. Kippenberg, Parametric normal-mode
	splitting in cavity optomechanics, Phys. Rev. Lett. \textbf{101}, 263602
	(2008).
	
	\bibitem{liu2013dynamic}
	Y.~C. Liu, Y.~F. Xiao, X.~Luan, and C.~W. Wong, Dynamic dissipative cooling of
	a mechanical resonator in strong coupling optomechanics, Phys. Rev. Lett.
	\textbf{110}, 153606 (2013).
	
	\bibitem{vitali2007optomechanical}
	D.~Vitali, S.~Gigan, A.~Ferreira, H.~B{\"o}hm, P.~Tombesi, A.~Guerreiro,
	V.~Vedral, A.~Zeilinger, and M.~Aspelmeyer, Optomechanical entanglement
	between a movable mirror and a cavity field, Phys. Rev. Lett. \textbf{98},
	030405 (2007).
	
	\bibitem{ghobadi2014optomechanical}
	R.~Ghobadi, S.~Kumar, B.~Pepper, D.~Bouwmeester, A.~Lvovsky, and C.~Simon,
	Optomechanical micro-macro entanglement, Phys. Rev. Lett. \textbf{112},
	080503 (2014).
	
	\bibitem{bai2019engineering}
	C.~H. Bai, D.~Y. Wang, S.~Zhang, S.~Liu, and H.~F. Wang, Engineering of strong
	mechanical squeezing via the joint effect between duffing nonlinearity and
	parametric pump driving, Photonics Res. \textbf{7}, 1229--1239 (2019).
	
	\bibitem{purdy2013strong}
	T.~P. Purdy, P.~L. Yu, R.~Peterson, N.~Kampel, and C.~Regal, Strong
	optomechanical squeezing of light, Phys. Rev. X \textbf{3}, 031012 (2013).
	
    \bibitem{bai2019qubit}
	C.~H. Bai, D.~Y. Wang, S.~Zhang, and H.~F. Wang, Qubit-assisted squeezing of
	mirror motion in a dissipative cavity optomechanical system, Sci. China Phys. Mech. Astron \textbf{62}, 970311 (2019).
	
	\bibitem{nunnenkamp2010cooling}
	A.~Nunnenkamp, K.~B{\o}rkje, J.~Harris, and S.~Girvin, Cooling and squeezing
	via quadratic optomechanical coupling, Phys. Rev. A \textbf{82}, 021806
	(2010).
	
	\bibitem{wang2018optomechanical}
	D.~Y. Wang, C.~H. Bai, S.~Liu, S.~Zhang, and H.~F. Wang, Optomechanical cooling
	beyond the quantum backaction limit with frequency modulation, Phys. Rev. A
	\textbf{98}, 023816 (2018).
	
	\bibitem{heinrich2011collective}
	G.~Heinrich, M.~Ludwig, J.~Qian, B.~Kubala, and F.~Marquardt, Collective
	dynamics in optomechanical arrays, Phys. Rev. Lett. \textbf{107}, 043603
	(2011).
	
	\bibitem{ludwig2013quantum}
	M.~Ludwig and F.~Marquardt, Quantum many-body dynamics in optomechanical
	arrays, Phys. Rev. Lett. \textbf{111}, 073603 (2013).
	
	\bibitem{xuereb2012strong}
	A.~Xuereb, C.~Genes, and A.~Dantan, Strong coupling and long-range collective
	interactions in optomechanical arrays, Phys. Rev. Lett. \textbf{109}, 223601
	(2012).
	
	\bibitem{akram2012photon}
	U.~Akram, W.~Munro, K.~Nemoto, and G.~Milburn, Photon-phonon entanglement in
	coupled optomechanical arrays, Phys. Rev. A \textbf{86}, 042306 (2012).
	
	\bibitem{xiong2015asymmetric}
	H.~Xiong, L.~G. Si, X.~Yang, and Y.~Wu, Asymmetric optical transmission in an
	optomechanical array, Appl. Phys. Lett. \textbf{107}, 091116 (2015).
	
	\bibitem{qi2019bosonic}
	L.~Qi, Y.~Yan, G.~L. Wang, S.~Zhang, and H.~F.~Wang, Bosonic Kitaev phase in a frequency-modulated optomechanical array, Phys. Rev. A \textbf{100}, 062323 (2019).	
	
	\bibitem{tomadin2012reservoir}
	A.~Tomadin, S.~Diehl, M.~D. Lukin, P.~Rabl, and P.~Zoller, Reservoir
	engineering and dynamical phase transitions in optomechanical arrays, Phys.
	Rev. A \textbf{86}, 033821 (2012).
	
	\bibitem{gan2016solitons}
	J.~H. Gan, H.~Xiong, L.~G. Si, X.~Y. L{\"u}, and Y.~Wu, Solitons in
	optomechanical arrays, Opt. Lett. \textbf{41}, 2676--2679 (2016).
	
	
	
		
	\bibitem{qi2017simulating}
	L.~Qi, Y.~Xing, H.~F. Wang, A.~D. Zhu, and S.~Zhang, Simulating $Z_{2}$ topological
	insulators via a one-dimensional cavity optomechanical cells array, Opt.
	Express \textbf{25}, 17948--17959 (2017).
	
	\bibitem{Roque2017Anderson}
	T. F. Roque, V. Peano, O. M. Yevtushenko, and F. Marquardt, Anderson localization of composite excitations in disordered optomechanical arrays. New J. Phys. \textbf{19}, 013006 (2017).
	
	\bibitem{Raeisi2020Quench}
	S. Raeisi and F. Marquardt, Quench dynamics in one-dimensional optomechanical arrays, Phys. Rev. A \textbf{101}, 023814 (2020).
	
	
	
	
	
	\bibitem{obana2019topological}
	D.~Obana, F.~Liu, and K. Wakabayashi, Topological edge states in the Su-Schrieffer-Heeger model, Phys. Rev. B
	\textbf{100}, 075437 (2019).
	
	\bibitem{liu2018topological}
	T.~Liu and H.~Guo, Topological phase transition in the quasiperiodic disordered Su–Schriffer–Heeger chain, Phys. Lett. A
	\textbf{382}, 3287 (2018).
	
	\bibitem{vos1996su}
	F. L. J. Vos, D. P. Aalberts, and W. van Saarloos, Su-Schrieffer-Heeger model applied to chains of finite length, Phys. Rev. B
	\textbf{53}, 14922 (1996).
	
	
	\bibitem{barone1977floquet}
	S. R. Barone, M. A. Narcowich, and F. J. Narcowich, Floquet theory and applications, Phys. Rev. A \textbf{15}, 1109 (1977).
	
	\bibitem{Berdanier2017floquet}
	W. Berdanier, M. Kolodrubetz, R. Vasseur, and J. E. Moore, Floquet Dynamics of Boundary-Driven Systems at Criticality, Phys. Rev. Lett. \textbf{118}, 260602 (2017).
	
\end{thebibliography}

\end{document}